\documentclass[12pt, draftclsnofoot, onecolumn]{IEEEtran}
\IEEEoverridecommandlockouts
\usepackage{verbatim}%for the comment
\usepackage[utf8x]{inputenc}
\usepackage{epstopdf}
\usepackage[pdftex]{graphicx}
\usepackage{tabularx}
\usepackage{booktabs}
\usepackage[table]{xcolor}
\usepackage{booktabs}
\usepackage{ctable}
\usepackage[cmex10]{amsmath}
\usepackage{amsthm}
\usepackage{algorithm}
\usepackage{algpseudocode}
\usepackage{pifont}
\usepackage{amssymb}
\usepackage{multirow}
\usepackage{cite}
\usepackage{textcomp}
\usepackage{array}
\usepackage{caption}
\usepackage{subcaption}
\graphicspath{{./Figures/}}

\theoremstyle{lemma}
\newtheorem{theorem}{Theorem}
\newtheorem{proposition}{Proposition}

\newcommand*\xbar[1]{%
	\hbox{%
		\vbox{%
			\hrule height 0.5pt % The actual bar
			\kern0.5ex%         % Distance between bar and symbol
			\hbox{%
				\kern-0.1em%      % Shortening on the left side
				\ensuremath{#1}%
				\kern-0.1em%      % Shortening on the right side
			}%
		}%
	}%
}

\newcommand{\U}{{\operatorname{U}}}
\newcommand{\M}{{\operatorname{M}}}
\newcommand{\Cl}{{\operatorname{Cl}}}
\newcommand{\B}{{\operatorname{B}}}
\newcommand{\D}{{\operatorname{D}}}
\newcommand{\G}{{\operatorname{G}}}
\newcommand{\A}{{\operatorname{A}}}

\begin{document}
	\bstctlcite{IEEEexample:BSTcontrol}
	%---------------------------------------------------------------------------------
	%                                    Title
	%---------------------------------------------------------------------------------
	\title{Energy-Efficient Massive IoT Shared Spectrum Access over UAV-enabled Cellular Networks} %The title of the paper
	\author{Ghaith Hattab,~\IEEEmembership{Student Member,~IEEE,} Danijela Cabric,~\IEEEmembership{Senior Member,~IEEE}
		\thanks{This paper was presented in part at the IEEE Int. Conf. on Wireless and Mobile Computing (WiMoB) \cite{Hattab2017a}. 	
			G. Hattab and D. Cabric are with the Department of Electrical and Computer Engineering, University of California, Los Angeles, CA 90095-1594 USA (email: ghattab@ucla.edu, danijela@ee.ucla.edu).}}

	\maketitle
	
	%---------------------------------------------------------------------------------
	%                                    Abstract
	%---------------------------------------------------------------------------------
	\vspace{-0.58in}\begin{abstract}
		Data aggregation has become an emerging paradigm to support massive Internet-of-things (IoT), a new and critical use case for fifth-generation new radio (5G-NR). Indeed, data aggregators can complement cellular base stations and process IoT traffic to reduce network congestion. In this paper, we consider using mobile data aggregators, e.g., drones, that collect IoT traffic and aggregate them to the network. Specifically, we first discuss how the spectrum can be shared between cellular users (UEs) and IoT devices in the presence of drones, proposing a time-division duplexing protocol. We use stochastic geometry to analyze this protocol, comparing it to the standard spectrum sharing and orthogonal allocation protocols. We then formulate a stochastic optimization problem to optimize the nominal IoT transmit power, maximizing the average energy-efficiency (EE) of the IoT device subject to interference constraints to protect UEs. Simulations are presented to validate the theoretical insights and the effectiveness of the proposed protocol. It is shown that using drones, to aggregate IoT traffic, improves the EE of IoT devices, yet the EE degrades as their altitudes increases. Equally important, optimizing the transmit power is critical to further improve the EE, while ensuring fair coexistence with UEs.
		\vspace{-0.1in}
	\end{abstract}
	
	%---------------------------------------------------------------------------------
	%                                    Index Words
	%---------------------------------------------------------------------------------
	\begin{IEEEkeywords} %For index terms
		Cellular networks, massive IoT, spectrum sharing, stochastic geometry, UAVs.	\vspace{-0.2in}
	\end{IEEEkeywords}

	%---------------------------------------------------------------------------------
	%                             I. Introduction
	%---------------------------------------------------------------------------------
	\section{Introduction}
	Fifth-generation new-radio (5G-NR) is set to unlock new application scenarios on different fronts. One specific use case is the native support of a massive number of sensors and machines, collectively known as massive cellular Internet-of-things (IoT) or massive machine-type communications (mMTC) \cite{ITUR2015}. Indeed, it is envisaged that billions of IoT devices will require Internet-connectivity by 2020, creating transformative economic potentials for operators and stakeholders\cite{Mckinsey2015} and spanning several vertical sectors such as smart cities and agriculture \cite{DawyYaacoub2017}. 
	
	Cellular networks have started to support new user categories tailored for IoT applications, e.g., NB-IoT and NR-lite, in addition to enabling extended discontinuous reception to reduce power consumption \cite{3GPP2015a,Rico-AlvarinoYavuz2016}. Such IoT optimizations, however, are still limited to services that do not require a large deployment of sensors and machines. Indeed, as the density of IoT devices increases, several challenges emerge \cite{Soltanmohammadi2016}. First, collisions among IoT devices increase due to the increased number of access requests, making retransmissions more frequent, and thus affecting their energy-efficiency (EE). Second, interference increases when IoT devices share the same spectrum with cellular users (UE), degrading the coverage needed for the former and reducing the spectral efficiency (SE) for the latter. In this paper, we propose a transmission protocol that uses drones to aggregate IoT traffic, while ensuring fair shared spectrum access with existing UEs.
	
	%------------
	%A. Related work
	%------------
	\subsection{Related work}
	The techniques toward the coexistence of IoT devices and UEs can be broadly classified into orthogonal-based and sharing-based solutions \cite{Rico-AlvarinoYavuz2016,Soltanmohammadi2016, Wali2017,Feng2017}. In the former, resource blocks are split among IoT devices and UEs as means to avoid interference and congestion. However, resource partitioning inevitably leads to a spectral efficiency tradeoff between IoT devices and UEs. In contrast, in spectrum sharing, all resource blocks are shared between UEs and IoT devices. To control congestion, several techniques have emerged such as access class barring (ACB) and randomized back-off schemes \cite{Jiang2018a,WangWong2015}. These methods, nevertheless, do not address the co-channel interference after access requests are granted. Alternative to these approaches, data aggregation has emerged as an effective solution to handle the massive IoT traffic. An experimental study is discussed in \cite{DawyYaacoub2017} showing the benefits of IoT data aggregation on cellular networks. In \cite{KwonCioffi2013,MalakAndrews2016,Tefek2017,Guo2017}, stochastic geometry is used to  analyze the coverage performance and/or the energy consumption using single and/or multiple aggregators. These works, however, consider fixed terrestrial aggregators and focus only on the performance of IoT devices, i.e., the coexistence of IoT devices and UEs is not considered. In this paper, we consider using unmanned aerial vehicles (UAVs) or drones that stop at optimized predetermined locations.
	
	Integrating UAVs with cellular networks has attracted significant attention, e.g., it has become a study item for recent and future 3GPP releases \cite{3GPP2017e}. Indeed, UAVs are envisioned to become data aggregators (relays) or even base stations (BSs) \cite{Bor-Yaliniz2016}, as their mobility brings unparalleled flexibility to cellular networks. For example, UAVs can help realize several IoT applications, e.g., smart cities, by extending the coverage of existing cellular infrastructure, enabling low-power communications with low-cost sensors, and reducing network congestion via offloading some of the traffic generated from massive IoT devices. Implementation and practical considerations for UAV-based IoT platforms are presented in \cite{Motlagh2017,Yuan2018}, showing the feasibility of using UAVs for IoT applications. Optimization and analysis of UAV-based aggregation are studied in \cite{Zeng2016,Bushnaq2018,Zhan2018,Mozaffari2017a, Koyuncu2018, Guo2019}. For example, the authors in \cite{Zeng2016} focus on optimizing the throughput of a single link between a source and a destination, assisted by a relaying UAV, whereas in our work we consider a large-scale cellular network. In \cite{Bushnaq2018}, the authors focus on minimizing the time it takes the drone to aggregate data samples collected by IoT devices to estimate a field of interest. In \cite{Zhan2018}, the authors study optimizing the UAV's trajectory and sensors' wake-up schedules to minimize their energy consumption. In \cite{Mozaffari2017a}, the locations and associations of the drones are optimized to minimize the transmit powers of IoT devices. In \cite{Koyuncu2018, Guo2019}, the authors optimize the deployment of UAVs to minimize the \emph{average nominal} transmit power of ground users. Compared to the aforementioned works, we mainly focus on the coexistence of UEs and IoT devices, using UAVs to enable a fair shared spectrum access. We further optimize the average nominal transmit power of IoT devices to maximize their EE.
	
	%------------
	%B. Contributions
	%------------
	\subsection{Contributions}
	The main contributions in this paper are as follows. 
	\begin{itemize}
		\item \emph{Shared spectrum-based transmission protocol}: We present a time-division duplexing (TDD) transmission protocol that provides a shared-spectrum access between massive IoT and UEs over the cellular network using UAVs as data aggregators for IoT devices. We use stochastic geometry \cite{Haenggi2012} to characterize the average available resources for UEs and IoT devices and compare the proposed protocol with a sharing-based protocol via ACB as well as resource splitting via frequency partitioning. We also analyze the coverage of the UEs in the presence of IoT devices transmitting to their associated drones. 
		\item \emph{Energy-efficiency maximization}: We then optimize the nominal transmit power of the IoT device to maximize its average EE subject to a protection criterion to its nearest UE. For tractable analysis, we first consider a single-cell single-drone (SC-SD) scenario, where the average EE is derived in closed form. We further analyze the interference-to-signal ratio (ISR) at the UE in the same cell with IoT devices and use its distribution as a protection constraint. Convexity analysis and insights on the optimal transmit power are discussed. We also present extensions to the problem, where the BS power is optimized and multiple drones per cell are considered.
	\end{itemize}
	We validate the theoretical expression of the EE and the effectiveness of optimizing the nominal transmit power of IoT devices via Monte Carlo simulations, showing that the proposed scheme provides significant EE improvements to IoT devices compared to transmitting at the maximum power, which is typically done to extend coverage \cite{3GPP2015a}. The proposed scheme is further compared to ACB and orthogonal allocation in a large network. Simulations show that the EE is significantly improved for practical drone altitudes, with minimal degradation to the UE's spectral efficiency in the UL and the DL. Such improvements are translated into an increased lifetime of IoT devices, which is validated using the 3GPP evaluation methodology in \cite{3GPP2015e}.
	
	%---------------------------------------------------------------------------------
	%                II. System Model
	%---------------------------------------------------------------------------------
	\section{System Model}\label{sec:system}
	
	%------------
	%A. Cellular network model
	%------------
	\subsection{Cellular network model}
	We use stochastic geometry to model the cellular network since it provides tractable analysis of large-scale networks \cite{Haenggi2012}. Such analysis helps understand the impact of different network parameters, gleaning useful design guidelines.
	
	%-------
	%1. BSs and UEs
	%-------
	\subsubsection{BSs and UEs}
	In stochastic geometry, the locations of BSs are commonly modeled using the homogeneous Poisson Point process (HPPP) $\Phi_\B$ with density $\lambda_\B$ \cite{Guo2017}. Each BS is equipped with a multi-antenna array with $M_\B$ antennas. We assume the BS can multiplex $1\leq U_\B \leq M_\B$ users per resource block. During downlink (DL) data communications, the BS transmits at a nominal power of $P_\B$, such that each multiplexed user is equally allocated a power of $P_\B/U_\B$.\footnote{In practice, the BS may optimize power allocation to UEs, e.g., using a water-filing algorithm that considers the BS-UE channel quality. Note if there are a large number of UEs in the network, it is reasonable to assume that the BS can find a subset of UEs with good channels, and so it can schedule them together and use equal power allocation.}
	For UEs, we assume that they are also generated from an independent HPPP $\Phi_\U$ with density $\lambda_\U$, where each one is equipped with a single-antenna system and connects to the nearest BS. During uplink (UL) data communications, the UE transmits at nominal power of $P_\U$.\footnote{We use the term nominal power here because, in general, the BS may request the UE to add an offset to the transmit power depending on the UE's estimated path loss. This scheme is known as fractional power control. The optimization of UE transmit power is outside the scope of this work.} We remark that we focus on a single-tier network for easier exposition in the subsequent analysis. However, it is straightforward to extend this work to multi-tier heterogeneous networks as we have done in our prior work in \cite{Hattab2017a}.
	
	%-------
	%2. Ground-to-ground channel model
	%-------
	\subsubsection{Ground-to-ground channel model}
	For ground-to-ground links, e.g., BS-UE links, we assume a power-law path loss model with a loss of $L_0$ at a reference distance of 1m and a decaying exponent of $\alpha_\G$. For small-scale fading, we assume a Rayleigh fading channel, with gamma distributed channel power gains, as they model a variety of multi-antenna transmission modes \cite{AfifyAlouini2016}. Specifically, the channel power gains between the UE and the tagged BS and the UE and an interfering BS are, respectively, modeled as $g_\B\sim\Gamma(\Delta_\B,1)$ and $f_\B\sim\Gamma(\Psi_\B,1)$, e.g., for multi-user zero-forcing beamforming, we have    $\Delta_\B=M_\B-U_\B+1$ and $\Psi_\B=U_\B$ \cite{AfifyAlouini2016}.

	%------------
	%B. IoT devices model
	%------------
	\subsection{IoT devices model} 
	We primarily consider UL IoT connectivity, which is typically the bottleneck in massive IoT communications \cite{DawyYaacoub2017}. Furthermore, we assume that IoT devices are clustered either inherently, e.g., deploying sensors in hotspots to monitor the same physical phenomenon \cite{Guo2017}, or via a clustering algorithm, as done in \cite{Mozaffari2017a}. To this end, we model the locations of IoT devices using an independent clustered HPPP process. In particular and similar to \cite{Guo2017}, we consider the Matern cluster process, where the locations of cluster centers, i.e., parent points, is modeled by the independent HPPP $\Phi_\Cl$ with density $\lambda_\Cl$. In each cluster, IoT devices represent the daughter points of the clustered process, denoted by $\Phi_\M$, and they are uniformly distributed in a disk of radius $R$ and with density $\lambda_\M$. Thus, the average density of IoT devices in the network is $\lambda_\M\lambda_\Cl$. Finally, we assume all IoT devices are single-antenna transmitters, and they transmit at a fixed power of $P_\M$. 
	
	Applications where such IoT models are reasonable include the mass deployment of IoT devices across a city, where sensors can be anchored on bridges for infrastructure monitoring, on buildings for utility metering, or in a farm for water management \cite{DawyYaacoub2017}. Such applications are delay tolerant, yet they require reliable coverage and a very long lifetime.

	%------------
	%C. Using UAVs for as data aggregators
	%------------
	\subsection{Using UAVs as data aggregators}
	
	We consider deploying UAVs, e.g., drones, as a middle layer between IoT devices and the cellular infrastructure. The drone acts as a \emph{mobile data aggregator} that is sent by a BS to provide coverage for a cluster of IoT devices. We note that while drones can be used as BSs or relays for UE traffic, as discussed in \cite{Bor-Yaliniz2016}, in this work we primarily use them as aggregators for IoT devices that have delay-tolerant traffic. We assume the density of drones, $\lambda_\D$, is equal to the density of clusters, i.e., $\lambda_\D=\lambda_\Cl$, and each one flies at an altitude of $h_\D$. We note that fewer drones can be also used, e.g., a drone can serve multiple clusters by moving from one point to another over time. Further, the drone is equipped with an omni-directional single-antenna cellular transceiver. We discuss the case of a multi-tier drone network, where each tier is defined by a different altitude, in  Section \ref{sec:general}.
	
	%-------
	%1. Initial access phase
	%-------
	\subsubsection{Initial access phase}
	Since we focus on massive IoT applications with machines and sensors anchored to fixed locations, it is reasonable to assume that the locations of IoT devices in the cluster are registered in a server or a database, and hence they are known to the mobile operator \cite{Mozaffari2017a,DawyYaacoub2017}. In particular, the drone, which is sent by the BS, moves to predetermined \emph{stop points} for UL data aggregation. In this work and for tractable analysis, the stop point of the $l$-th drone, $(x_{\D,l},y_{\D,l},h_\D)$,  is the centroid of its cluster of IoT devices. We note that such location minimizes the distance to the typical IoT device, i.e., it can be shown that $(x_{\D,l},y_{\D,l})= \operatorname*{argmin}_{(x,y)} \mathbb{E}_{\Phi_\M}[d_{\M,l}]$, where $d_{\M,l}$ is the 3D distance between a typical IoT device and the drone. Finally, we note that the trajectory of the drone and its mobility can be optimized to extend its lifetime as discussed in \cite{Zeng2017, Koyuncu2018}. In this work, it suffices to assume that drones can manage to hover around a cluster of IoT devices to collect data from them.
	
	%-------
	%2. Ground-to-air channel model
	%-------
	\subsubsection{Ground-to-air channel model}
	We consider the following popular ground-to-air path loss model for the link between an IoT device and the $l$-th drone \cite{Mozaffari2016, Mozaffari2017a,MozaffariDebbah2016a,Al-HouraniLardner2014}
	\begin{equation}
	\begin{aligned}
	l_{\M\rightarrow\D}(d_{\M,l})&= \mathbb{P}_{\operatorname{LOS}}(d_{\M,l},h_\D) L_0 d_{\M,l}^{-\alpha_{\A}}
	&+  (1-\mathbb{P}_{\operatorname{LOS}}(d_{\M,l},h_\D) ) L_{\operatorname{NLOS}} L_0 d_{\M,l}^{-\alpha_{\A}},
	\end{aligned}
	\end{equation}
	where $\mathbb{P}_{\operatorname{LOS}}(d_{\M,l},h_\D)$ is the line-of-sight (LOS) probability, $ L_{\operatorname{NLOS}}$ is the excessive path loss due to non-LOS, and $\alpha_{\A}$ is the ground-to-air path loss exponent. Note that, as investigated in \cite{Al-HouraniLardner2014}, the impact of multi-path fading is negligible in such links, and thus it is ignored. The LOS probability is found using the 3GPP UMa-AV channel model, and it can be expressed as \cite{3GPP2017e}
	\begin{equation}
	\label{eq:LOSprobability}
	\mathbb{P}_{\operatorname{LOS}}(d_{\M,l},h_\D)  = \operatorname{min}\left\{\frac{\xi_1}{r_{\M,l}},1\right\}\left(1-e^{-r_{\M,l}/\xi_2}\right) + e^{-r_\M/\xi_2},
	\end{equation} 
	where $r_{\M,l} = \displaystyle \sqrt{d_{\M,l}^2-h_\D^2}$, and $\xi_1$ and $\xi_2$ are some constants as given in \cite[Table B-1]{3GPP2017e}. For the link between the BS and the drone, we consider a similar model, which is expressed as 
	\begin{equation}
	\begin{aligned}
	l_{\B\rightarrow\D}(d_{\B,l})&= L_{\operatorname{STR}}\cdot \Big(\mathbb{P}_{\operatorname{LOS}}(d_{\B,l},h_\D) L_0 d_{\B,l}^{-\alpha_{\A}} &+  (1-\mathbb{P}_{\operatorname{LOS}}(d_{\B,l},h_\D) ) L_{\operatorname{NLOS}} L_0 d_{\B,l}^{-\alpha_{\A}}\Big),
	\end{aligned}
	\end{equation}
	where the attenuation $L_{\operatorname{STR}}$ is due to the fact that the BS's antenna array steering direction typically points horizontally or is tilted downwards when communicating with ground users \cite{3GPP2017a}. Such attenuation depends on the drone's relative location with the array's boresight. However, we assume it to be constant for tractable analysis, which is reasonable when the drone is flying at heights higher than the BS's antenna height. 
	
\begin{figure}[t!]
		\centering
		\begin{subfigure}[t]{.45\textwidth}
			\centering
			\includegraphics[width=3.0in]{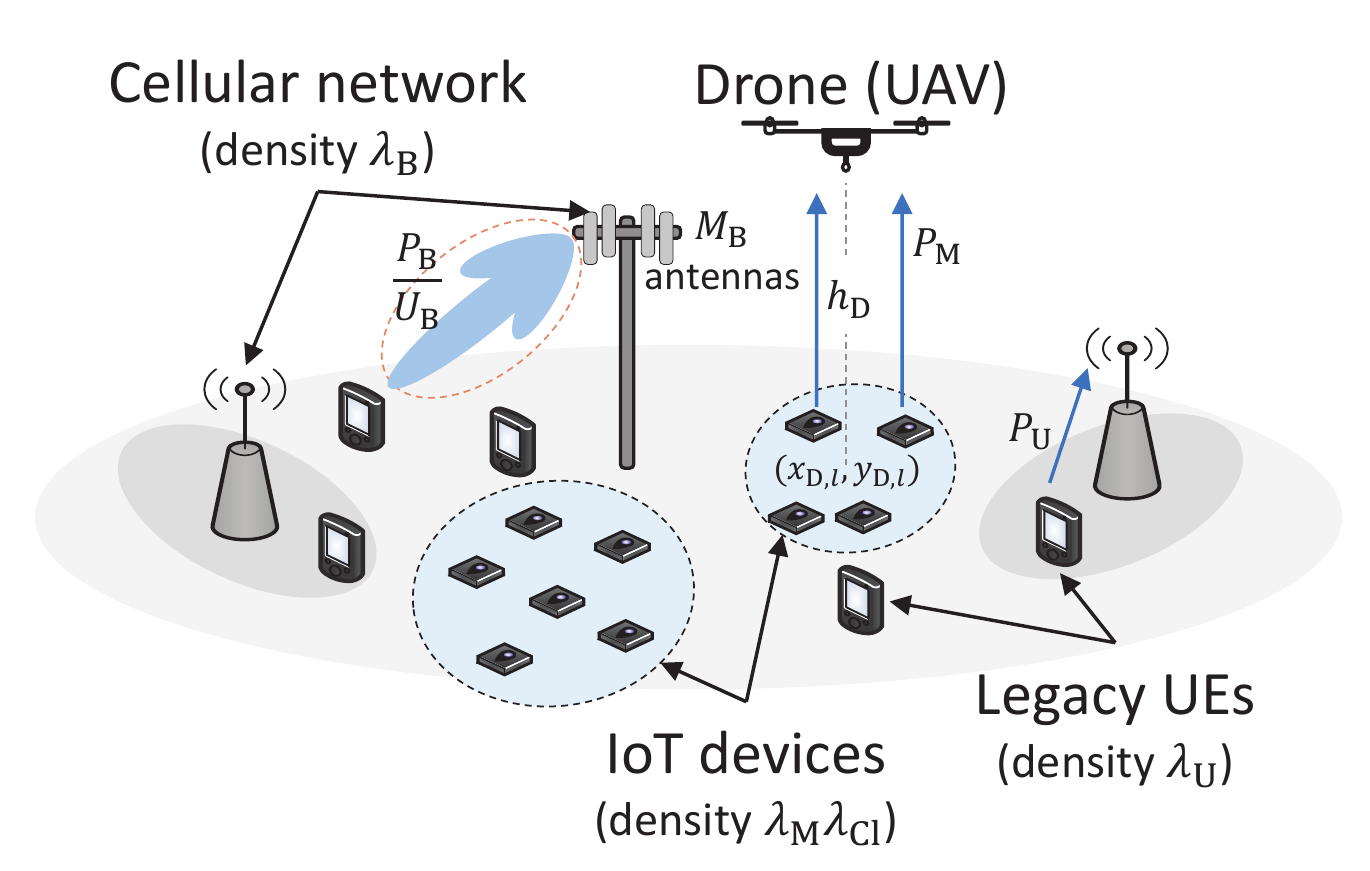}
			\caption{}
			\label{fig:architecture}
		\end{subfigure}
		\begin{subfigure}[t]{.45\textwidth}
			\centering
			\includegraphics[width=3.0in]{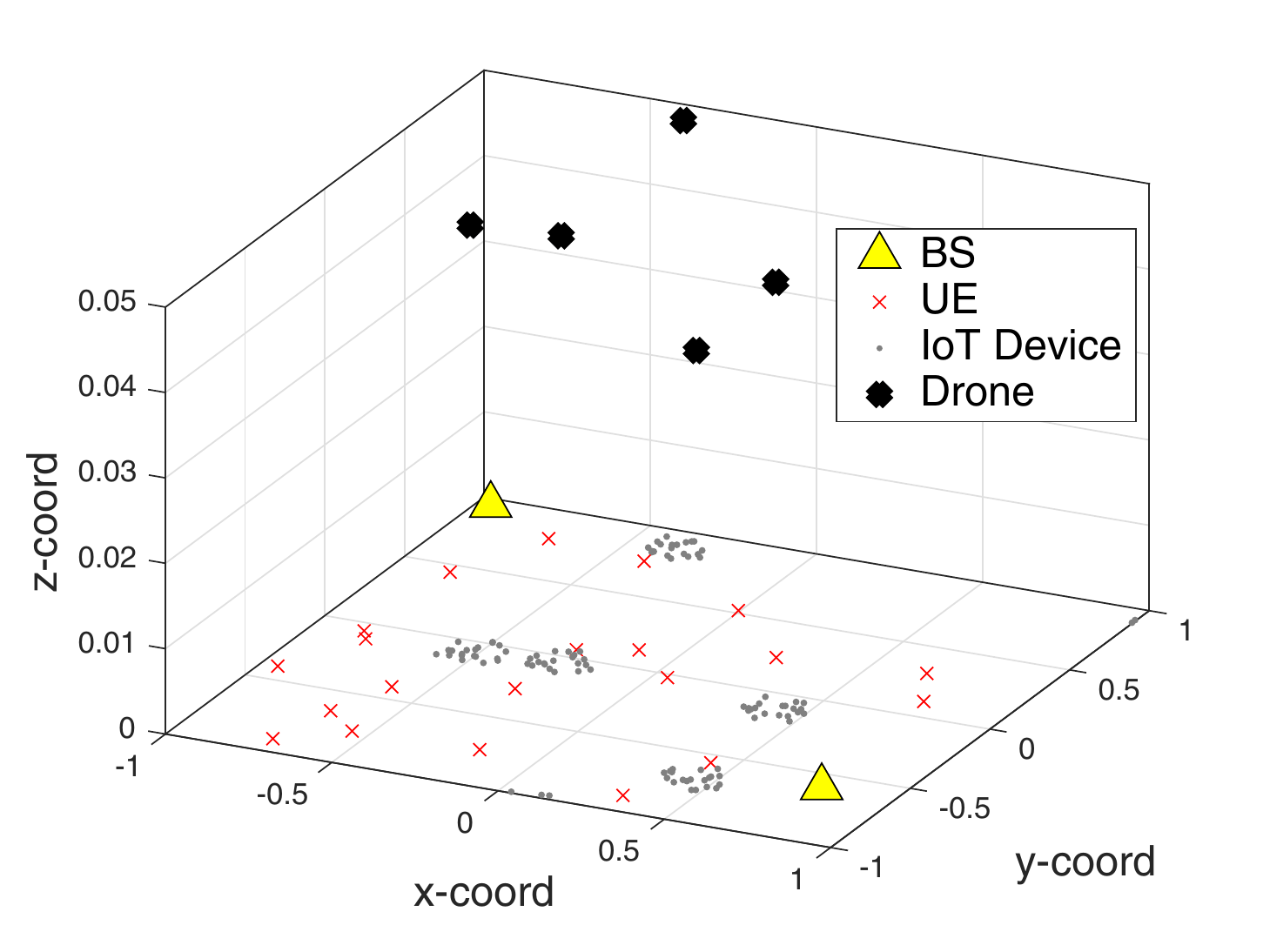}
			\caption{}
			\label{fig:spatialRealization}
		\end{subfigure}
		\caption{(a) The UAV-enabled cellular architecture; (b) A spatial realization of the network topology ($\lambda_\U=5\lambda_\B$, $\lambda_{\D}=\lambda_\B$, and $\lambda_{\M}=20$/cluster).}
		\label{fig:system}
		\vspace{-.20in}
	\end{figure}

	The UAV-enabled cellular network is shown in Fig. \ref{fig:architecture}, where a drone is sent by a BS to a cluster of IoT devices, stops at $(x_{\D,l},y_{\D,l})$ to collect data from IoT devices, and then aggregates the traffic to the BS. An illustration of one realization of the network is also shown in Fig. \ref{fig:spatialRealization}.
	
	%------------
	%D. Performance Metrics
	%------------
	\subsection{Performance Metrics}
	We consider two metrics: the spectral efficiency (SE) of a typical UE in the UL and DL and the energy efficiency (EE) of a typical IoT device, which are defined next.
	
	%-------
	%1. UE Spectral Efficiency
	%-------
	\subsubsection{UE Spectral Efficiency} 
	We consider the UE to employ a multi-modulation and coding scheme, and thus the spectral efficiency of a typical UE in the DL/UL is defined as \cite{BaoLiang2015}
	\begin{equation}
	\label{eq:SE}
	C_{\U,\xi} = \bar \beta_{\U,\xi} \sum_{k=1}^{K_\U} \log_2(1+\tau_{\U,k}) \boldsymbol{1}(\tau_{\U,k+1}\geq \gamma_{\U,\xi}\geq \tau_{\U,k}),
	\end{equation}
	where $\xi\in\{\operatorname{DL},\operatorname{UL}\}$,  $\gamma_{\U,\xi}$ is the signal-to-interference-plus-noise ratio (SINR), $\{\tau_{\U,k}\}$ are the SINR thresholds,  $\boldsymbol{1}(\cdot)$ is the indicator function, and $ \bar \beta_{\U,\xi}$ is a pre-log term that denotes the UE long-term available resources in time, frequency, and space. This rate model assumes $K_\U$ possible thresholds, and it is a generalization of the single-rate model, i.e., $K_\U=1$, that is commonly used in the literature \cite{LinLiang2015,AfifyAlouini2016}. We emphasize that the mean load approximation, i.e., the assumption that the load $ \bar \beta_{\U,\xi}$ is independent of $\gamma_{\U,\xi}$ \cite{BaoLiang2015,LinLiang2015}, is used only for tractable analysis, while we relax this assumption when we run Monte Carlo simulations in Section \ref{sec:simulations}.
	
	%-------
	%2. IoT Energy Efficiency
	%-------
	\subsubsection{IoT Energy Efficiency} 
	We consider the EE of the IoT device, i.e., the ratio of the achieved rate, $r(P_\M)$, to the total power consumption, $c(P_\M)$. More formally, the EE of a typical IoT device is defined as \cite{ZapponeDebbah2016}
	\begin{equation}
	\label{eq:EE}
	\begin{aligned}
	E_\M \triangleq \frac{r(P_\M)}{c(P_\M)} = \frac{\bar \beta_{\M}  \sum_{k=1}^{K_\M} \log_2(1+\tau_{\M,k}) \boldsymbol{1}(\tau_{\M,k+1}\geq \gamma_\M\geq \tau_{\M,k})}{P_{\operatorname{CP}}+\eta^{-1}P_{\M}},
	\end{aligned}
	\end{equation}
	where $P_{\operatorname{CP}}$ is a constant that quantifies the circuit power consumption, $\eta$ is the power amplifier efficiency, and $\bar \beta_\M$ and $\gamma_\M$ are the IoT device long-term resources and SINR, respectively. 
	The motivation behind using this particular metric is as follows. We aim to address two conflicting objectives: maximizing the rate and minimizing the transmit power. Thus, one approach is to formulate a \emph{scalarized} multi-objective optimization problem \cite{BjornsonOttersten2014}, where the objective function is a weighted sum of $r(P_\M)$ and $c(P_\M)$. In such a problem, different weights lead to different \emph{Pareto} optimal solutions, i.e., operating points that lie on the \emph{Pareto boundary}, where improving one objective value can only degrade the other objective value. It can be shown that the ratio of the two objectives, in this case the EE, is one of the points on the Pareto boundary \cite{BjornsonOttersten2014}. The ratio here also has a physical interpretation, i.e., the efficiency of the communication protocol measured in the output bps per unit power consumed. Moreover, the metric is also relevant to applications where coverage is paramount. For example, the value of $\tau_{\M,1}$ determines the minimum coverage needed since a zero rate, and hence zero EE, is achieved if $\gamma_\M<\tau_{\M,1}$. Third, the metric is also applicable to devices that only support a single modulation scheme, where we would set $K_{\M}=1$.
	
	We note that in this paper, we focus on optimizing the link between the IoT device and the drone. The link between the drone and the BS can be optimized separately, e.g., the drone can get closer to the BS to ensure reliable aggregation \cite{Mei2018}, the drone can compress data generated from devices performing a similar task \cite{Biason2017}, or the drone can itself be a BS \cite{Bor-Yaliniz2016}. Since this link is studied in the aforementioned works, it is outside the scope of this paper. A summary of the main parameters is given in Table \ref{tab:parameters}.  
	
	\begin{table*}[!t]
		\caption{Main parameters and their values if applicable}
		\tiny
		\renewcommand{\arraystretch}{1.2}
		\label{tab:parameters}
		\centering
		\begin{tabular}{|c|l|l|}
			\hline
			\textbf{Description }  				&  \textbf{Parameters} &Value(s) (if applicable)\\\hline
			\multirow{4}{*}{Path loss parameters} 			&$\alpha_\G$: Path loss exponent for ground-to-ground links &$\alpha_\G=3.5$ \\\cline{2-3}
			&$\alpha_\A$: Path loss exponent for ground-to-air links &$\alpha_\A=2.2$ \cite{3GPP2017e} \\\cline{2-3}
			&$L_0$: Path loss at a reference distance of 1m (assuming 2GHz carrier frequency)&$L_0=-38$dB\\\cline{2-3}
			&$L_{\operatorname{NLOS}}$: Additional non-LOS path loss &$L_{\operatorname{NLOS}}=-20$dB \cite{MozaffariDebbah2016a}\\\cline{2-3}
			&$L_{\operatorname{STR}}$: Additional path loss due to BS's antenna array steering direction &$L_{\operatorname{STR}}=-30$dB \cite{3GPP2017e}  \\\cline{1-3}		
			\multirow{3}{*}{BS parameters}  	& $P_\B$: BS transmit power & $P_\B=\{46,32\}$dBm \\\cline{2-3}
			&$M_\B$: Number of antennas at the BS& $M_\B=32$\\\cline{2-3}
			&$U_\B$: Number of spatially multiplexed users & $U_\B=4$ \\\cline{1-3}
			\multirow{2}{*}{UE parameters}  	& $P_\U$: UE transmit power & $P_\U=23$dBm \\\cline{2-3}
			&$\tau_{\U,k}$: SINR threshold & from $-5$ to $30$dB\\\cline{1-3}
			\multirow{5}{*}{IoT device parameters}& $P_{\M}^{\operatorname{max}}$ and $P_{\M}^{\operatorname{min}}$: Maximum and minimum allowable transmit powers & $P_{\M}^{\operatorname{max}}=23$dBm and  $P_{\M}^{\operatorname{min}}=1$dBm\\\cline{2-3}
			&$P_{\operatorname{CP}}$: Circuit power consumption & $P_{\operatorname{CP}}=90$mW\cite{3GPP2015e}\\\cline{2-3}
			&$\eta$: PA efficiency &$\eta=0.44$ \cite{3GPP2015e}\\\cline{2-3}
			&$\tau_{\M,k}$: SINR threshold &from $-5$ to $10$dB \\\cline{2-3}
			&$R$: Cluster radius & $R=50$m \cite{Guo2017}\\\hline
		\end{tabular}
		\vspace{-0.1in}
	\end{table*}
	
	%---------------------------------------------------------------------------------
	%                         III. TDD protocol for shared spectrum access
	%---------------------------------------------------------------------------------
	\section{TDD protocol for shared spectrum access}\label{sec:TDD}
	In this section, we present the proposed transmission protocol to enable shared spectrum access between massive IoT and UEs over UAV-enabled cellular networks. We then analyze the average allocated resources of the UE and the IoT device under the proposed protocol and compare it to those achieved under existing transmission protocols. 
	
\begin{figure*}[t!]
		\centering
		\begin{subfigure}[t]{.32\textwidth}
			\centering
			\includegraphics[width=2.25in]{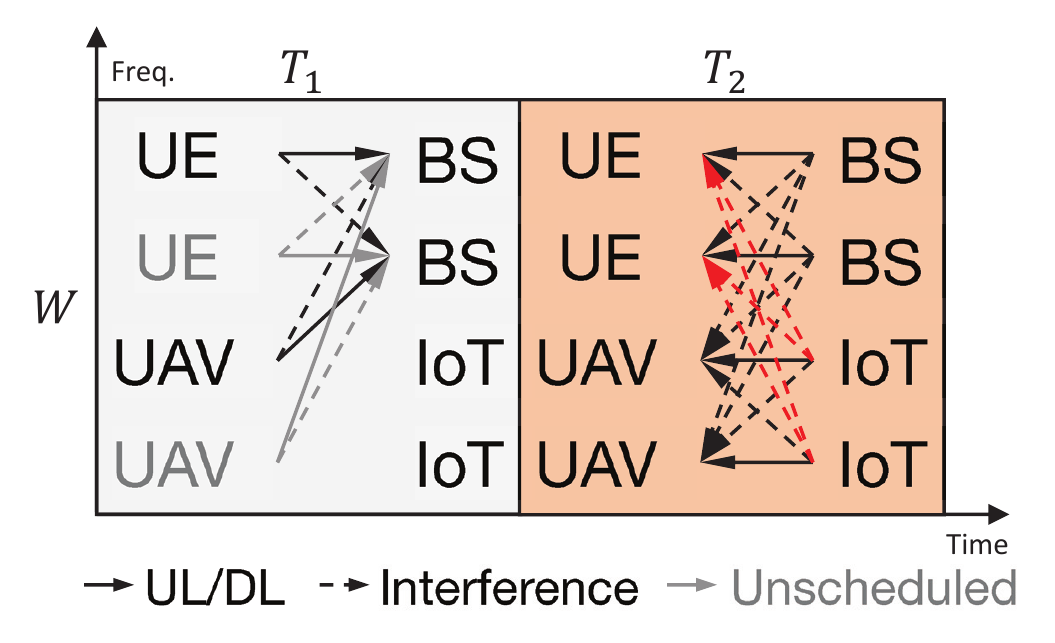}
			\caption{\small Proposed TDD}
			\label{fig:proposedTDD}
		\end{subfigure}	
		\begin{subfigure}[t]{.32\textwidth}
			\centering
			\includegraphics[width=2.25in]{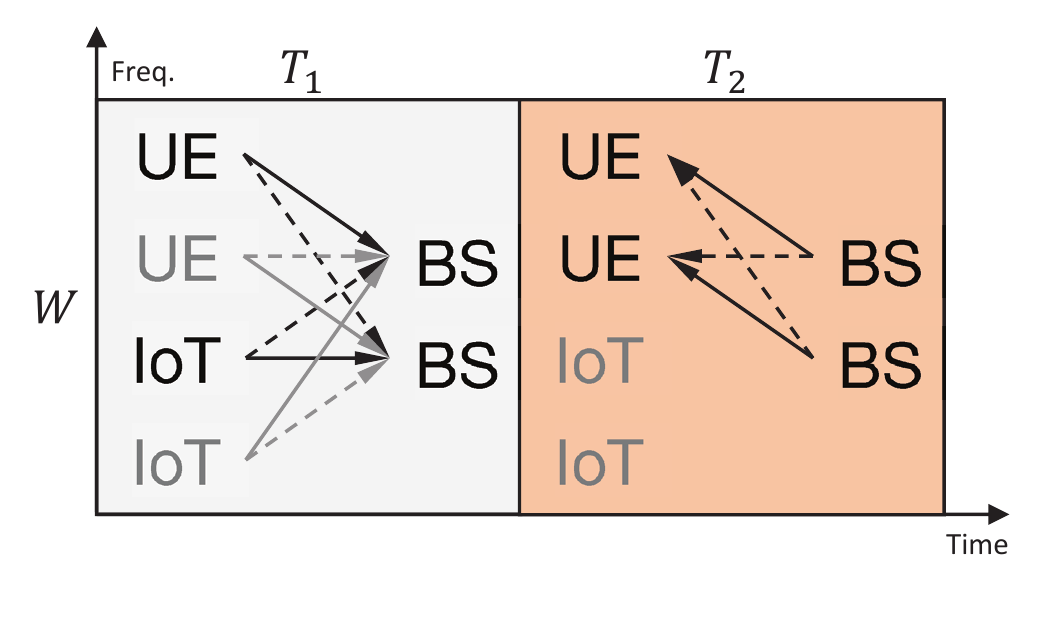}
			\caption{\small Sharing-based TDD}
			\label{fig:sharingTDD}
		\end{subfigure}
		\begin{subfigure}[t]{.32\textwidth}
			\centering
			\includegraphics[width=2.25in]{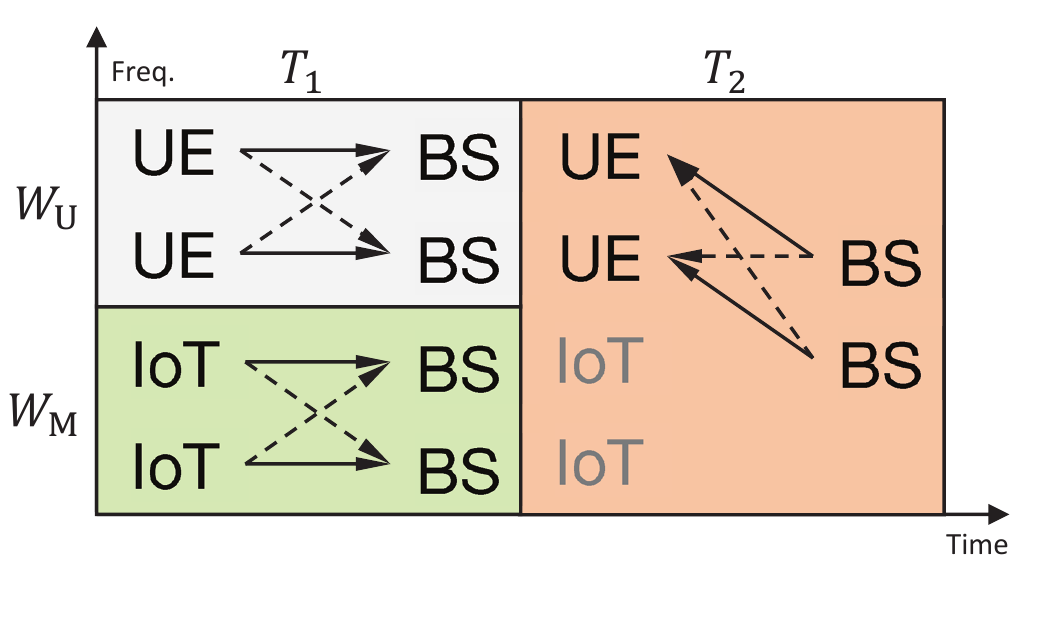}
			\caption{\small Orthogonal-based TDD}
			\label{fig:orthogTDD}
		\end{subfigure}
		\caption{Comparison of the different TDD transmission protocols.}
		\label{fig:protocols}
		\vspace{-.15in}
	\end{figure*}

	We focus on TDD cellular networks, and thus the proposed protocol is divided into two slots: $T_1$ and $T_2$. In the first time slot, the UE operates in the UL, communicating with its tagged BS. Similarly, in this slot, we treat the drone as another UE, which aggregates previously collected data from IoT devices and sends it to its tagged BS. In the second time slot, the UE operates in the DL, whereas the IoT device operates in the UL, communicating with its associated drone, as shown in Fig. \ref{fig:proposedTDD}. The proposed protocol is motivated as follows.\footnote{The proposed protocol differs from the reverse TDD protocol in \cite{SanguinettiDebbah2015}. The latter is developed to address the coexistence of macro and small cells in the presence of wireless backhaul. In addition, in a given time slot, not all cells use the same spectrum in reverese TDD, and further, small cells are divided into two groups. The first small cell group operates in the UL, i.e., from small cell to macro cell, and the second one operates in DL, i.e., from small cell to UE.}
	\begin{itemize}
		\item \textbf{Maximum bandwidth}: The protocol allows the IoT device to share the same time-frequency block with the UE, i.e., no resource splitting is used. 
		\item \textbf{Reducing congestion}: When the UE operates in the UL, it competes for scheduling with drones instead of IoT devices, and thus the channel congestion is significantly reduced. 
		\item \textbf{Limiting the impact of IoT interference}: The shared access paradigm inevitability leads to additional interference from IoT devices into UEs. However, when IoT devices transmit data in the UL, UEs operate in the DL,  where their tagged BSs transmit at much higher power than the transmit power of IoT devices as $P_\B\gg P_\M$.
	\end{itemize}
	
	%------------
	%A. Characterization of average resources in the proposed and existing protocols
	%------------
	\subsection{Characterization of average resources in the proposed and existing protocols}\label{sec:loadAnalysis}
	For tractable analysis, we assume that a proportional fair scheduler is used, e.g., round-robin, and thus under the mean load approximation \cite{LinLiang2015}, the average load is inversely proportional to the average number of devices connected to the same source. 
	
	%-------
	%1. Proposed protocol
	%-------
	\subsubsection{Proposed protocol} 
	Let $\bar \beta_{\U,\operatorname{UL}}^{\operatorname{P}}$ denote the average allocated resources to a typical UE operating in the UL under the proposed protocol. Then, it can be shown that
	\begin{equation}
	\begin{aligned}
	\bar \beta_{\U,\operatorname{UL}}^{\operatorname{P}} &\stackrel{(a)}{=} \underset{\operatorname{Freq.}}{\underbrace{W}} \times \underset{\operatorname{Space}}{\underbrace{U_\B}}  \times  \underset{\operatorname{Time}}{\underbrace{T_1 \left(\frac{1}{N_{\B,\operatorname{UL}}^{\operatorname{P}}}\right)}} \\
	&\stackrel{(b)}{=} W \times U_\B \times T_1 \left(\frac{\lambda_\B}{\lambda_\U+\lambda_\D}\right),
	\end{aligned}
	\end{equation}
	where $(a)$ follows since the entire bandwidth $W$ is allocated to the UE, $U_\B$ UEs can be spatially multiplexed by the same BS, and the portion of the time allocated to the UE is inversely proportional to the average number of devices connected to the BS, i.e., $N_{\B,\operatorname{UL}}^{\operatorname{P}}$. Here, $(b)$ follows by showing that the average number of UEs (or drones) connected to the BS under the nearest BS association is $\lambda_\U/\lambda_\B$ (or $\lambda_\D/\lambda_\B$) \cite{LinLiang2015}. Similarly, the average portion of resources of the UE in the DL is expressed as $
	\bar \beta_{\U,\operatorname{DL}}^{\operatorname{P}}= W \times U_\B \times T_2 \left(\frac{\lambda_\B}{\lambda_\U}\right)$,
	which follows since in the second time slot, no IoT devices are connected to the BS under the proposed protocol. For the IoT device, it can be shown that the average portion of resources is given as $\bar \beta_\M^{\operatorname{P}} = W \times T_2\lambda_\M^{-1}$, which follows since the average number of IoT devices per cluster is $\lambda_\M$, and the drone multiplexes one IoT device per time-frequency slot. Note that for low-rate applications, the drone can divide the bandwidth $W$ into smaller frequency blocks and serve multiple IoT devices, one per frequency block. This does not change $\bar \beta_\M^{\operatorname{P}}$ since the decrease in frequency resources is equally compensated by increased time resources.  
	
	%-------
	%2. Comparison with sharing-based protocol
	%-------
	\subsubsection{Comparison with sharing-based protocol}
	In the sharing-based protocol, the IoT device is registered as another UE, as shown in Fig. \ref{fig:sharingTDD}. To control channel congestion, the 3GPP standard proposes one mechanism, namely \emph{access class barring} (ACB), which prioritizes UE traffic over IoT devices \cite{Soltanmohammadi2016}. More formally, each BS broadcasts a parameter $0\leq \kappa \leq 1$ to all IoT devices in vicinity. The IoT device then generates a random number $n\in[0,1]$ before initiating a channel access request. If  $n>\kappa$, then the IoT device does not request access. Clearly, for $\kappa=1$, the protocol simplifies to a standard sharing protocol that is agnostic to the device type. Let $\bar \beta_{\U,\operatorname{UL}}^{\operatorname{S}}$ denote the average allocated resources to a typical UE operating in the UL under the sharing-based protocol. Then, we have $
	\bar \beta_{\U,\operatorname{UL}}^{\operatorname{S}} = W \times U_\B \times T_1  \left(\frac{\lambda_\B}{\lambda_\U+\kappa \lambda_\M \lambda_\Cl}\right)$,
	which follows since the average number of IoT devices per BS is $ \lambda_\M \lambda_\Cl/\lambda_\B$, yet only a fraction, $\kappa$, of them request access. Clearly, for $\bar \beta_{\U,\operatorname{UL}}^{\operatorname{S}}>\bar \beta_{\U,\operatorname{UL}}^{\operatorname{P}}$, we must have $\kappa<\lambda_\M^{-1}$, yet this degrades the average resources allocated to the IoT device. Indeed, the mean resources for the IoT device under the sharing protocol is $\bar \beta_\M^{\operatorname{S}} =\mathbb{P}(n>\kappa) \bar \beta_{\M|n>\kappa}^{\operatorname{S}} +  \mathbb{P}(n\leq\kappa) \bar \beta_{\M|n>\kappa}^{\operatorname{S}}$, which can be simplified to 
	\begin{equation}
	\begin{aligned}
	\bar \beta_\M^{\operatorname{S}} &=   W \times U_\B \times \kappa T_1  \left(\frac{\lambda_\B}{\lambda_\U+\kappa \lambda_\M \lambda_\Cl}\right)~
	\stackrel{(a)}{\leq} \bar \beta_\M^{\operatorname{P}} \left(\frac{ U_\B\lambda_\B}{\lambda_\U+\lambda_\Cl}\right),
	\end{aligned}
	\end{equation} 
	where $(a)$ follows using $\kappa<\lambda_\M^{-1}$ and assuming $T_1=T_2$. Since the density of UEs is typically higher than the density of BSs, i.e., $\lambda_\U\gg \U_\B\lambda_\B$, we get $\bar \beta_\M^{\operatorname{S}}<\bar \beta_\M^{\operatorname{P}}$. Finally, for the second time slot, we have $\bar \beta_{\U,\operatorname{DL}}^{\operatorname{S}} =\bar \beta_{\U,\operatorname{DL}}^{\operatorname{P}}$ for the UE. 
	
	%-------
	%3. Comparison with orthogonal-based protocol
	%-------
	\subsubsection{Comparison with orthogonal-based protocol}
	Alternative to using ACB to resolve channel congestion, 3GPP has also proposed the separation of frequency resources \cite{Soltanmohammadi2016}, as shown in Fig. \ref{fig:orthogTDD}. Let $W_\U\in[0,W]$ be the portion of the spectrum allocated for UEs, then the average portion of allocated resources for a typical UE in the UL under this protocol is $
	\bar\beta_{\U,\operatorname{UL}}^{\operatorname{O}} = W_\U \times U_\B \times T_1 \left(\frac{\lambda_\B}{\lambda_\U}\right)$.
	If $W_\U/W> \lambda_\U/(\lambda_\U+\lambda_\D)$, then we have $\bar\beta_{\U,\operatorname{UL}}^{\operatorname{O}}>\bar\beta_{\U,\operatorname{UL}}^{\operatorname{P}}$. However, this comes at the expense of reducing the resources allocated to the IoT device since we have $\bar\beta_\M^{\operatorname{O}} = W_\M \times U_\B \times T_1 \left(\frac{\lambda_\B}{\lambda_\M\lambda_\Cl}\right)$, which is bounded by 
	\begin{equation}
	\begin{aligned}
	\bar\beta_\M^{\operatorname{O}} %&= W_\M \times U_\B \times T_1 \left(\frac{\lambda_\B}{\lambda_\M\lambda_\Cl}\right)\\
	&\stackrel{(a)}{<}  W \times U_\B \times \frac{T_1}{\lambda_\M}  \left(\frac{\lambda_\B}{\lambda_\U+\lambda_\Cl}\right)\\
	&\stackrel{(b)}{<} \bar\beta_\M^{\operatorname{P}},
	\end{aligned}
	\end{equation} 
	where $(a)$ follows from the fact that $W_\M=W-W_\U$ and thus $
	\frac{W_\U}{W} > \frac{\lambda_\U}{\lambda_{\U}+\lambda_\Cl}
	\stackrel{}{\Longleftrightarrow} \frac{W_\M}{W} < \frac{\lambda_\Cl}{\lambda_{\U}+\lambda_\Cl}$. 
	In addition, $(b)$ follows assuming $T_1=T_2$ and the density of UEs is high. Finally and similar to the sharing-based protocol, we have $\bar \beta_{\U,\operatorname{DL}}^{\operatorname{O}} =\bar \beta_{\U,\operatorname{DL}}^{\operatorname{P}}$ for the UE. 
	
	To summarize, for the sharing protocol to outperform the proposed one in terms of resource allocation for the UE, we must use aggressive ACB with very small values of $\kappa$, which inevitably affects the IoT EE as the average portion of allocated resources is decreased. For the orthogonal allocation to outperform the proposed protocol in terms of the UE performance, nearly all resources should be allocated to the UE, i.e., $W_\U\rightarrow W$, since $\lambda_\U\gg \lambda_\D$, and this also limits the resources allocated to the IoT device. We note that the derived theoretical allocated resources for the proposed protocol, i.e., $(\bar \beta_{\U,\operatorname{UL}}^{\operatorname{P}}, \bar \beta_{\U,\operatorname{DL}}^{\operatorname{P}}, \bar \beta_{\M}^{\operatorname{P}})$, are agnostic to the type of aggregator used. Indeed, using drones versus terrestrial aggregators primarily affect the signal and interference powers and not the number of resources allocated to devices.
	
	%------------
	%B. Analysis of IoT Interference on UEs
	%------------
	\subsection{Analysis of IoT Interference on UEs}
	While the proposed protocol improves the average allocated resources of a typical UE, in comparison with existing protocols, the signal-to-interference ratio (SIR) of a typical UE degrades due to the presence of an additional interference term generated from IoT devices (cf. Fig. \ref{fig:proposedTDD}). To study the coverage probability of the typical UE, we define the UE SIR as
	\begin{equation}
	\label{eq:UESINR}
	\tilde \gamma_{\U,\operatorname{DL}}^{\operatorname{P}} = \frac{\frac{P_\B}{U_\B}g_\B L_0 x_\B^{-\alpha_\G}}{\sum_{y_b\in\Phi_\B'} \frac{P_\B}{U_\B}f_b L_0 y_b^{-\alpha_\G}+\sum_{z_m\in\Phi_\M'} P_\M f_m L_0 z_m^{-\alpha_\G}},
	\end{equation}
	where $\Phi_\B'$ is the set of interfering BSs, $\Phi_\M'$ is the set of interfering IoT devices, $f_m\sim\Gamma(1,1)$, $x_\B$ is the distance to the tagged BS, $y_b$ is the distance to the $b$-th interfering BS, and $z_m$ is the distance to the $m$-th interfering IoT device. %Next, we derive the coverage probability of the UE under the proposed protocol to highlight the key parameters that affect the UE's coverage. 
	
	The coverage probability is defined as $\mathbb{C}^{\operatorname{P}}_{\U,\operatorname{DL}}(\tau)	\triangleq \mathbb{P}(\tilde \gamma_{\U,\operatorname{DL}}^{\operatorname{P}}\geq \tau)$, which can be rewritten as 
	\begin{equation}
	\begin{aligned}
	\mathbb{C}^{\operatorname{P}}_{\U,\operatorname{DL}}(\tau)
	&\stackrel{(a)}{=} 2\pi \lambda_\B \int_{0}^{\infty} x \mathbb{E}_{g_\B}\left[F_{I_\U}\left(\frac{g_\B}{\tau x^{\alpha_\G}}\right)\right] e^{-\pi \lambda_\B x^2}dx,
	\end{aligned}
	\end{equation}
	where $(a)$ follows from the distribution of the distance from the UE to the tagged BS \cite{LinLiang2015} and $F_{I_\U}(\cdot)$ is the cumulative distribution function (CDF) of the  interference. Using the Gil-Pelaez Inversion theorem \cite{GilPelaez1951} to compute the interference CDF, we get the following theorem. 
	
	\begin{theorem}\label{theo:UEcoverage}
		The coverage probability of the UE under the proposed protocol is expressed as 
		\begin{equation}
		\label{eq:coverage}
		\mathbb{C}^{\operatorname{P}}_{\U,\operatorname{DL}}(\tau)= \frac{1}{2} - \frac{\Upsilon(\tau,\lambda_\B,\lambda_\D,\hat P_{\B,\M})}{\pi}, 
		\end{equation}
		where $\hat P_{\B,\M}=\operatorname{sinc}^{-1}(\delta_\G)P_\M U_\B/P_\B$, $\delta_\G=2/\alpha_\G$, and
		\begin{equation}
		\begin{array}{ll}
		\Upsilon(\tau,\lambda_\B,\lambda_\D,\hat P_{\B,\M}) =  \int_{0}^{\infty}\frac{1}{t}\operatorname{Im}\left\{\frac{(1+jt/\tau)^{-\Delta_\B}}{{}_2F_1(\delta_\G,\Psi_\B;1-\delta_\G;jt)+\frac{\lambda_\D}{\lambda_\B}\hat P_{\B,\M}^{\delta_\G}(-jt)^{\delta_\G}}\right\}	dt.
		\end{array}
		\end{equation}
	\end{theorem}
	\emph{Proof:} See Appendix A. $\hfill\blacksquare$
	
	The expression is given in a single integral form that can be efficiently evaluated using numerical software. The key insight here is that the degradation of the UE coverage due to the presence of IoT devices depends mainly on two factors: (i) the ratio of the transmit power of the IoT device to the power allocated to the UE, i.e., $P_\M/(P_\B/U_\B)$ and (ii) the average number of drones per BS, i.e., $\lambda_\D/\lambda_\B$. To see this, note that the coverage probability in the absence of IoT devices is $
	\mathbb{C}^{\operatorname{N}}_{\U,\operatorname{DL}}(\tau)= \frac{1}{2} - \frac{1}{\pi}\int_{0}^{\infty}\frac{1}{t}\operatorname{Im}\left\{\frac{(1+jt/\tau)^{-\Delta_\B}}{{}_2F_1(\delta_\G,\Psi_\B;1-\delta_\G;jt)}\right\}dt$, 
	and thus using stochastic dominance, i.e., $\mathbb{C}^{\operatorname{N}}_{\U,\operatorname{DL}}(\tau)\geq \mathbb{C}^{\operatorname{P}}_{\U,\operatorname{DL}}(\tau)$, we have  $
	\int_{0}^{\infty}\frac{1}{t}\operatorname{Im}\left\{\frac{(1+jt/\tau)^{-\Delta_\B}}{{}_2F_1(\delta_\G,\Psi_\B;1-\delta_\G;jt)}\right\}	dt \leq  \Upsilon(\tau,\lambda_\B,\lambda_\D,\hat P_{\B,\M})$, 
	where the gap above decreases as $\lambda_\D/\lambda_\B\rightarrow0$ and/or  $P_\M/(P_\B/U_\B)\rightarrow0$. 
	
	\begin{figure}[t!]
		\centering
		\begin{subfigure}{.45\textwidth}
			\center
			\includegraphics[width=3.0in]{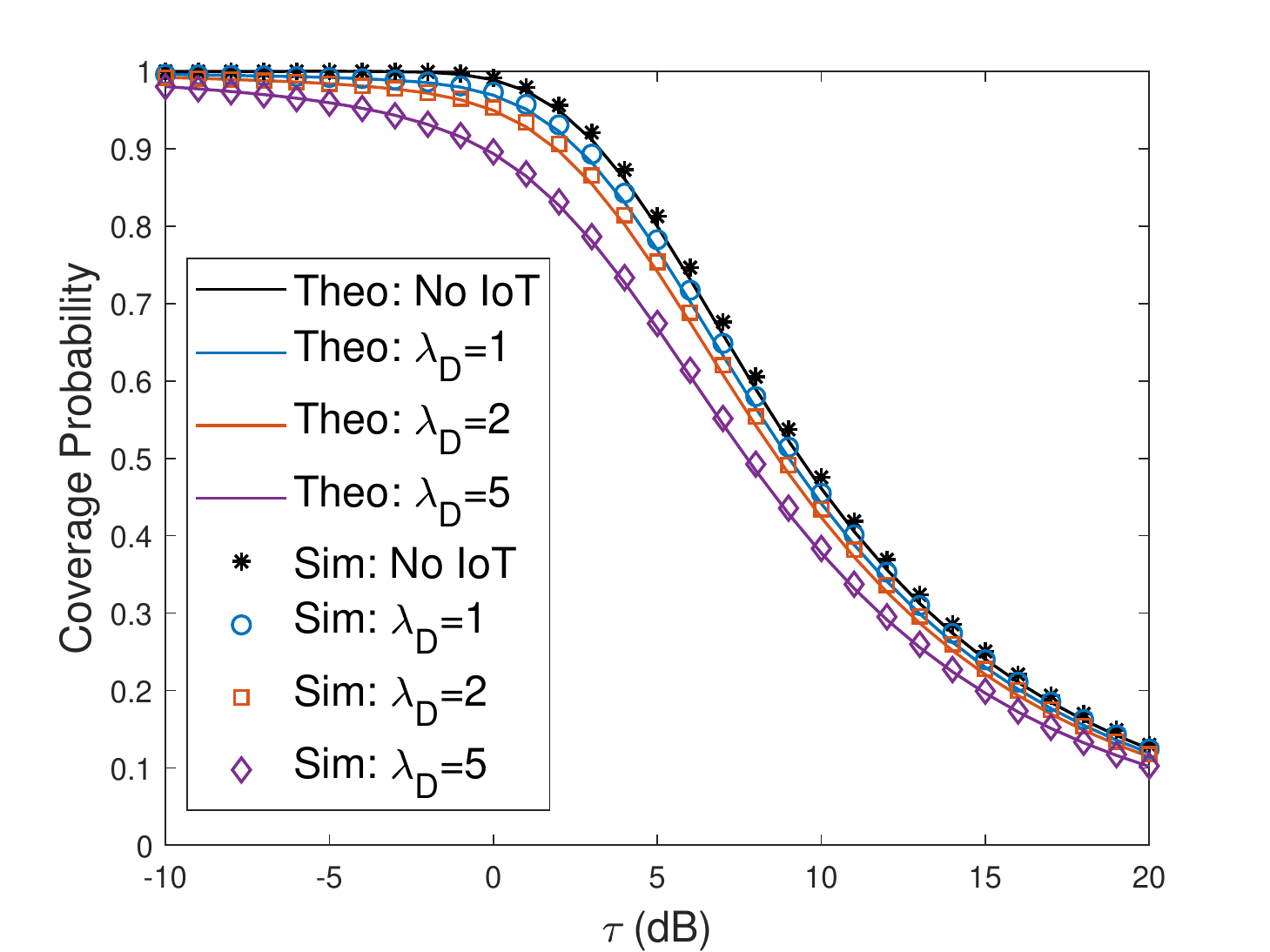}
			\caption{Variations of SIR threshold}
			\label{fig:UECov_tau}
		\end{subfigure}
		\begin{subfigure}{.45\textwidth}
			\center
			\includegraphics[width=3.0in]{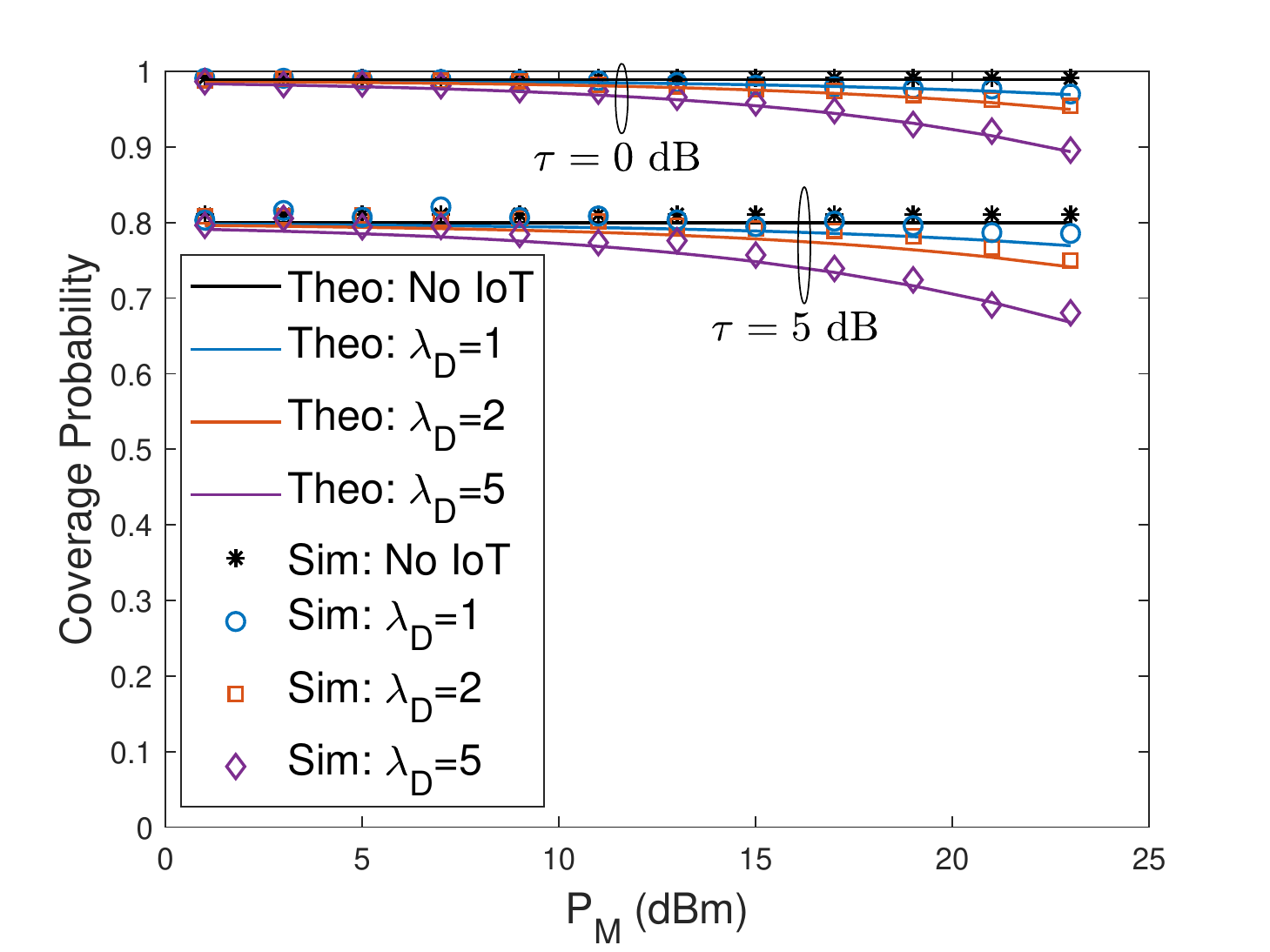}
			\caption{Variations of IoT transmit power, $P_\M$}
			\label{fig:UECov_Pm}
		\end{subfigure}
		\caption{The UE DL coverage performance with and without IoT devices.}
		\label{fig:PcovTwoTier}
		\vspace{-0.1in}
	\end{figure}

	We validate the theoretical expression with Monte Carlo simulations, using the same parameters in Table \ref{tab:parameters}. Fig. \ref{fig:UECov_tau} shows the distribution of coverage in the presence and absence of IoT devices. It is observed that increasing the number of drones decreases the coverage, yet the degradation is not severe, e.g., the median SIR merely degrades by 1.7dB under the proposed protocol with $\lambda_\D/\lambda_\B=5$. Recall that here a drone is assigned to a single cluster. However, if a single drone is assigned instead to serve multiple clusters, moving from one stop point to another, then the interference on UEs from IoT transmission decreases, e.g., median SIR degradation is less than 0.36dB when $\lambda_\D/\lambda_\B=1$. Nevertheless, the cost of reducing this interference is the increased delay on IoT devices, leading eventually to degradation of the IoT energy efficiency when $\lambda_\M\lambda_\Cl\gg1$. To this end, another approach to limit the interference is to reduce the IoT transmit power, as shown in Fig. \ref{fig:UECov_Pm}. In the next section, we focus on optimizing the IoT transmit power such that the IoT EE is maximized and the IoT interference is controlled.

	%---------------------------------------------------------------------------------
	%                         IV. IoT Energy-Efficiency Maximization
	%---------------------------------------------------------------------------------
	\section{IoT Energy-Efficiency Maximization}\label{sec:EEoptimization}
	In this section, we formulate a stochastic optimization problem to maximize the average EE of an IoT device under the proposed protocol. The problem has the following form
	\begin{equation}
	\label{eq:EEOpt}
	\begin{array}{cl}
	\underset{P_\M}{\text{{maximize}}}
	&~~ \mathbb{E}[E_\M] \\
	\text{subject to}     
	&~~f(I_\U)\leq \epsilon, \\
	&~P_\M\in\mathcal{P},\\
	\end{array}
	\end{equation}
	where the $f(\cdot)$ is an interference constraint to protect UEs and $\mathcal{P}$ is the feasible set of transmit powers. We have the following remarks about the problem in (\ref{eq:EEOpt}). First, it is a stochastic optimization framework, as the objective function is the mean of a random variable, and the expectation is taken with respect to spatial realizations. Second, such formulation aims to mainly optimize the nominal transmit power, $P_\M$, significantly reducing the complexity of implementation, i.e., the network or the drone only broadcasts the optimal value as a reference power over a control channel\footnote{In LTE and 5G-NR, the BS sends power control commands via the downlink control information (DCI) over the physical downlink control channel (PDCCH). In this case, the reference power, also known as open-loop transmit power, $P_o$, is set to the solution of the problem, i.e., $P_o = P_\M^\star$. Once the device successfully decodes the DCI, it follows the power control commands on the uplink channel, possibly adding offsets to $P_o$ depending on the path loss or channel quality.} to all of its IoT devices. Thus, all devices belonging to the same cluster use the same transmit power, which maximizes on \emph{average} the EE, i.e., this approach does not necessarily maximize the EE of every device as a single value is used, yet it significantly reduces the control overhead. Third, the constraint is a function of the interference from IoT devices into UEs. We do not enforce a UE rate constraint because this would incur additional overhead due to the necessary coordination between UEs and UAVs.
	There are two challenges to solve (\ref{eq:EEOpt}). The first one is that the objective function is intractable due to the ground-to-air channel model from interfering BSs and IoT devices into drones. The second one is that the UE coverage probability	in (\ref{eq:coverage}) is given in an integral form, making it not amenable to use as a UE protection criterion. For these reasons, we focus on optimizing the EE in a single-cell single-drone (SC-SD) setting, i.e., we ignore the interference from BSs and IoT devices outside the cell of the typical IoT device. We discuss several extensions to this problem in Section \ref{sec:general}.

	%------------
	%A. Average IoT EE in the SC-SD Case
	%------------
	\subsection{Optimization of the IoT EE in the SC-SD Case}
	
	\subsubsection{Derivation of the average EE}
	Let $\hat E_\M$ and $\hat \gamma_\M$ denote the IoT EE and IoT UL SINR in the SC-SD case, respectively, and let $p=P_\M$ for notational simplicity. Then, the average EE under the proposed protocol, i.e., $\bar E_\M(p) \triangleq \mathbb{E}[\hat E_\M]$, can be written as  $
	\bar E_\M(p) =  \frac{\bar \beta_{\M}^{\operatorname{P}} \sum_{k=1}^{K_\M} \mu_k \mathbb{P}(\hat \gamma_\M(p)\geq \tau_{\M,k})}{P_{\operatorname{CP}}+\eta^{-1}p}$, 
	where $\mu_1=\log_2(1+\tau_{\M,1})$ and $\mu_k=\log_2(1+\tau_{\M,k+1})-\log_2(1+\tau_{\M,k})$ for $k>1$. 
	
	In a single cell, the drone receives signals from IoT devices in its cluster and receives interference from the tagged BS, which operates in the DL. The next proposition presents the distributions of the received desired signal and interference powers at the drone, which will be useful to evaluate $\mathbb{P}(\hat \gamma_\M(p)\geq \tau_{\M,k})$. 
	
	\begin{proposition}\label{prop:IoTDistributions}
		The distribution of the desired signal power, $S_\M=P_\M l_{\M\rightarrow\D}(d_{\M,l})$, at a typical drone is expressed as 
		\begin{equation}
		\label{eq:IoTSignal}
		\mathbb{P}(S_\M\leq \tau)= 1 - \frac{(P_\M L_\M/\tau)^{\delta_\A}-h_\D^2}{R^2},
		\end{equation}
		where $\delta_\A=2/\alpha_\A$, $L_\M=L_0((1-L_{\operatorname{NLOS}}) \bar{\mathbb{P}}_{\operatorname{LOS},\M}+L_{\operatorname{NLOS}})$, $\bar{\mathbb{P}}_{\operatorname{LOS},\M}$ is the average of (\ref{eq:LOSprobability}), which is computed numerically,  and $\tau\in[P_\M L_\M (R^2+h_\D^2)^{-1/\delta_\A},P_\M L_\M h_\D^{-\alpha_\A}]$. In addition, the distribution of the interference signal power $I_\B=P_\B  l_{\B\rightarrow\D}(d_{\B,l})$ is given as
		\begin{equation}
		\label{eq:BSInterference}
		\mathbb{P}(I_\B\leq \tau)= \exp(\pi \lambda_\B h_\D^2)\cdot  \exp\left(-\pi \lambda_\B (P_\B L_\B/\tau)^{\delta_\A}\right),
		\end{equation}
		where $L_\B=L_0((1-L_{\operatorname{NLOS}}L_{\operatorname{STR}}) \bar{\mathbb{P}}_{\operatorname{LOS},\B}+L_{\operatorname{NLOS}}L_{\operatorname{STR}})$ and $\tau\in[0,P_\B L_\B h_\D^{-\alpha_\A}]$. 
	\end{proposition}
	\emph{Proof:} See Appendix B. $\hfill\blacksquare$
	
	\emph{Remark}: It is observed from (\ref{eq:IoTSignal}) that as $h_\D$ increases, the variations in $S_\M$ reduces since $\tau\in[P_\M L_\M (R^2+h_\D^2)^{-1/\delta_\A},P_\M L_\M h_\D^{-\alpha_\A}]$, i.e., the distribution of $S_\M$ becomes more concentrated around the median. This follows because the drone location is optimized to reduce the average 2D distance to a typical IoT device, increasing the LOS probability particularly when $h_\D\geq R$. Similarly, it is observed from (\ref{eq:BSInterference}) that increasing the density of BSs increases the interference as the average distance between the drone and its tagged BS decreases. Furthermore, it can be shown that increasing the drone's height decreases $I_\B$, but not as rapidly as the decrease in $S_\M$. 
	
	Using Proposition \ref{prop:IoTDistributions}, we derive the coverage probability and the average EE of the IoT device. 
	
	\begin{theorem}\label{theo:IoTEE}
		The coverage probability of an IoT device in the SC-SD case is given as
		\begin{equation}
		\label{eq:IoTCov}
		\mathbb{P}(\hat \gamma_\M\geq \tau) \approx e^{\pi \lambda_\B h_\D^2}\cdot  \exp\left(-\pi \lambda_\B \left(\frac{\frac{P_\M \tilde L_\M}{\tau} - P_N}{P_\B L_\B}\right)^{-\delta_\A}\right),
		\end{equation}
		where $\tilde L_\M=L_\M (\frac{R^2}{2}+h_{\D}^2)^{-1/\delta_\A}$ and $P_N$ is the noise power. In addition, the average EE of an IoT device is expressed as 
		\begin{equation}
		\label{eq:IoTEE}
		\begin{aligned}
		\bar E_\M(p) =  \frac{\bar \beta_{\M}^{\operatorname{P}}  e^{\pi \lambda_\B h_\D^2} \sum_{k=1}^{K_\M} \mu_k  e^{-\frac{\pi \lambda_\B}{(P_\B L_\B)^{-\delta_A}} \left(\frac{p \tilde L_\M}{\tau_k} - P_N\right)^{-\delta_\A}}}{P_{\operatorname{CP}}+\eta^{-1}p}.
		\end{aligned}
		\end{equation}
	\end{theorem} 
	\emph{Proof:} The coverage probability can be written as
	\begin{equation}
	\begin{aligned}
	\mathbb{P}(\hat \gamma_\M\geq \tau)&\triangleq  \mathbb{P}\left(\frac{S_\M}{I_\B+P_N}\geq \tau\right)
	&\approx  \mathbb{P}\left(I_\B\leq \frac{\tilde S_\M}{\tau} - P_N\right),
	\end{aligned}
	\end{equation}
	where we have used the median received signal power $\tilde S_\M$ computed from (\ref{eq:IoTSignal}) to approximate the coverage (recall that $S_\M$ has small variations particularly when $h_\D\geq R$). Thus, using (\ref{eq:BSInterference}), we arrive at (\ref{eq:IoTCov}). $\hfill\blacksquare$
	
	%------------
	%B. IoT Interference Constraint
	%------------
	\subsubsection{IoT Interference Constraint}
	We consider the distribution of the interference-to-signal ratio (ISR) as a protection criterion. We note that in \cite{Wei2016}, the mean of ISR (MISR) has been used as a metric to compare different interference mitigation techniques over cellular networks with stochastic topologies. Yet, using the ISR distribution provides more flexibility compared to the MISR, e.g., the protection criterion can be designed to limit the median, the 95th percentile, etc. The ISR is defined as $
	\operatorname{ISR}_\U \triangleq \frac{\mathbb{E}_{f_\M}[\hat I_{\U,\M}]}{\mathbb{E}_{g_\B}[S_\B]} = \frac{P_\M z_\M^{-\alpha_\G}}{\frac{\Delta_\B P_\B}{U_\B} z_\B^{-\alpha_\G}}$,
	where $\hat I_{\U,\M}$ is the interference from the typical IoT device to its nearest UE and $S_\B$ is the received desired signal power at the UE from its tagged BS. We note, similar to the MISR metric, the expectation is first taken with respect to channel realizations. The distribution is thus given as
	\begin{equation}
	\label{eq:IoTConstraint}
	\begin{aligned}
	\mathbb{P}(\operatorname{ISR}_\U\geq \rho)
	%&\stackrel{(a)}{=} \mathbb{P}\left(\frac{P_\M z_\M^{-\alpha_\G}}{\frac{\Delta_\B P_\B}{U_\B} z_\B^{-\alpha_\G}}\geq\rho\right)\\
	%&= \mathbb{P}\left(\frac{z_\B}{z_\M}\geq\left(\rho\frac{\Delta_\B P_\B}{U_\B P_\M} \right)^{1/\alpha_\G}\right)\\
	&\stackrel{(a)}{=} \mathbb{E}_{z_\M}\left[\exp\left(-\pi \lambda_\B z_\M^{2}\left(\rho\frac{\Delta_\B P_\B}{U_\B P_\M} \right)^{\delta_\G}\right) \right]\\
	&\stackrel{(b)}{=} \left[1+\frac{\lambda_\B}{\lambda_\U}\left(\rho\frac{\Delta_\B P_\B}{U_\B P_\M} \right)^{\delta_\G}\right]^{-1},
	\end{aligned}
	\end{equation}
	where $(a)$ follows using the complementary CDF of the distance from the UE to its tagged BS and $(b)$ follows by taking the expectation with respect to the distance between the IoT device and the nearest UE. 
	
	%------------
	%C. The SC-SD optimization problem
	%------------
	\subsubsection{The SC-SD problem}
	Using (\ref{eq:IoTEE}) as an objective function and the ISR expression in (\ref{eq:IoTConstraint}) as an interference constraint, i.e., $f(I_\U)\leq \epsilon \Leftrightarrow \mathbb{P}(\operatorname{ISR}_\U\geq \rho)\leq \epsilon$, we have
	\begin{equation}
	\label{eq:SCSDOpt}
	\begin{array}{cl}
	\underset{P_\M}{\text{{maximize}}}
	&~~ \frac{\sum_{k=1}^{K_\M} \mu_k  \exp\left(-\frac{\pi \lambda_\B}{(P_\B L_\B)^{-\delta_A}} \left(\frac{P_\M \tilde L_\M}{\tau_k} - P_N\right)^{-\delta_\A}\right)}{P_{\operatorname{CP}}+\eta^{-1}P_\M} \\
	\text{subject to}     
	&~~P_\M \leq \rho \left(\frac{\Delta_\B P_\B}{U_\B}\right) \left(\frac{\lambda_\B}{\lambda_\U}\cdot \frac{\epsilon}{1-\epsilon}\right)^{1/\delta_\G}, \\
	&~P_\M^{\operatorname{min}}\leq P_\M\leq P_\M^{\operatorname{max}},\\
	\end{array}
	\end{equation}
	where $P_\M^{\operatorname{min}}$ and $P_\M^{\operatorname{max}}$ are the minimum and maximum allowable transmit powers, respectively. In what follows, we denote the numerator of the objective function, i.e., the rate, by $r(P_\M)$, and the denominator, i.e., power consumption, by $c(P_\M)$. The following proposition shows that the problem in (\ref{eq:SCSDOpt}) is quasiconcave, and thus a local maximum is a global one \cite{BoydVandenberghe2004}. 
	
	\begin{proposition}\label{prop:SCSDQCNC}
		The objective function in (\ref{eq:SCSDOpt}) is quasiconcave and unimodal, whereas the constraints are all affine. Hence, the optimization problem is quasiconcave. 
	\end{proposition}
	\emph{Proof:} The proof follows from showing that $r(p)$ is S-shaped in $p$ (see Appendix C). $\hfill\blacksquare$
	
	Since the problem has a single optimizing variable, a line search is sufficient to solve the problem. A closed-form solution can be derived for the special case $K_\M=1$ and $\alpha_\A=2$. Specifically, the objective function, in this case, is maximized at 
	\begin{equation}
	\begin{aligned}
	\label{eq:OpTPowerSpecial}
	p^\star_{\operatorname{unconst}}&= \frac{\tau_{\M,1} (P_N+\pi \lambda_\B P_\B L_\B)}{\tilde L_\M}
	&+\frac{\sqrt{\pi \tau_{\M,1}\lambda_\B P_\B L_\B ( P_N \tau_{\M,1}+  \tilde L_M \eta P_{\operatorname{CP}})}}{ \tilde L_\M},
	\end{aligned}
	\end{equation}
	and thus the optimal transmit power is $P_\M^\star= \min\{P_\M^{\operatorname{max}},\rho (\frac{\Delta_\B P_\B}{U_\B}) \left(\frac{\lambda_\B}{\lambda_\U}\cdot \frac{\epsilon}{1-\epsilon}\right)^{1/\delta_\G},p^\star_{\operatorname{unconst}}\}$. It is observed from (\ref{eq:OpTPowerSpecial}) that the optimal power increases for: (i) higher target threshold $\tau_{\M,1}$ to meet the new coverage requirement, (ii) higher BS transmit power $P_\B$ or higher BS density $\lambda_\B$ to combat the increased interference from the cellular network, or (iii) higher PA efficiency $\eta$ to utilize the decrease in power consumption.

	%---------------------------------------------------------------------------------
	%Generalizations to the SC-SD Framework
	%---------------------------------------------------------------------------------
	\subsection{Generalizations to the SC-SD problem}\label{sec:general}
	
	%------------
	%A. Optimizing BS transmit power
	%%------------
	\subsubsection{Optimizing BS transmit power}
	The BS transmit power can be optimized with $P_\M$. Indeed, $P_\B$ affects both the objective function and the interference constraint in (\ref{eq:SCSDOpt}). To this end, if the optimized nominal transmit power $P_\M^\star$ satisfies the interference constraint with strict inequality, then $P_\B$ can be reduced so that the constraint is met with equality, reducing the interference seen at the drone. In return, $P_\M$ can be set lower due to the reduction in interference, and thus the EE is further improved. This procedure can be done recursively, as summarized in Alg. \ref{alg:BSopt}, until no further improvements are achieved. Here, $P_{\B}^{\operatorname{min}}$ and $P_{\B}^{\operatorname{max}}$ determine the range of the feasible BS transmit power.
	
	\begin{algorithm} [t!]
		\small
		\caption{Optimizing BS transmit power}\label{alg:BSopt}
		\begin{algorithmic}[1]
			\Procedure{($P_{\B,0}=P_{\B}^{\operatorname{max}}$; $\zeta>0$ )}{}
			\While{$\left|\bar E(P_{\M,i+1})-\bar E(P_{\M,i})\right|>\zeta$}
			\State \textbf{Update constraint:} $I_{\operatorname{const},i}=\rho \left(\frac{\Delta_\B P_{\B,i}}{U_\B}\right) \left(\frac{\lambda_\B}{\lambda_\U}\cdot \frac{\epsilon}{1-\epsilon}\right)^{1/\delta_\G}$
			\State \textbf{Solve for $P_{\M,i+1}$ in (\ref{eq:SCSDOpt}) using line search} 
			\State \textbf{Update BS transmit power:} $P_{\B,i+1}=\min\left(\max\left( \frac{P_{\M,i+1}}{\rho \frac{\Delta_\B}{U_\B} \left(\frac{\lambda_\B}{\lambda_\U}\cdot \frac{\epsilon}{1-\epsilon}\right)^{1/\delta_\G}},P_{\B}^{\operatorname{min}}\right),P_{\B}^{\operatorname{max}}\right)$
			\State $i=i+1$
			\EndWhile
			\State \textbf{return} $P_\M^\star=P_{\M,i}$ and $P_\B^\star=P_{\B,i}$
			\EndProcedure
		\end{algorithmic}
	\end{algorithm}

	%------------
	%B. EE Optimization in heterogeneous single-cell multi-drone case
	%------------
	\subsubsection{Generalization to the single-cell multi-drone case}\label{subsec:SCMD}
	We consider the single-cell multi-drone case (SC-MD), where each cell has multiple drones, each serving a cluster of IoT devices. We can further assume each drone belongs to a different tier, i.e., we consider an $N$-tier drone network such that the $l$-th tier drone flies at an altitude of $h_{\D,l}$ and serves an IoT cluster of radius $R_l$. The IoT device in a cluster served by the $l$-th drone transmits at power $P_{\M,l}$. 
	
	The challenge in the multi-drone is that the objective function can no longer be given in a closed-form expression due to the complicated ground-to-air path loss model between multiple interferers and the drone. Here, the interferers, with respect to a given drone, are the tagged BS and the other IoT devices transmitting to their respective drones in the same cell. To this end, we propose to model the different interference sources in the cell as one Poissonian source with a transmit power equal to the sum of transmit powers of all interferers. Such an approach ensures a tractable formulation and will be validated in the simulations section. Under this modeling assumption, the EE of an IoT device served by the $l$-th drone tier can be written as
	\begin{equation}
	\label{eq:EEobjSCMD}
	\bar E_{l}(\boldsymbol{P}_\M)\triangleq \frac{\bar \beta_{\M}^{\operatorname{P}}  e^{\pi \lambda_\B h_\D^2} \sum_{k=1}^{K_\M} \mu_k  e^{-\frac{\pi \lambda_\B \left(\frac{P_{\M,l} \tilde L_{\M,l}}{\tau_k} - P_N\right)^{-\delta_\A}}{(P_\B L_\B+\sum_{n\neq l}P_{\M,n} \tilde L_{\M,n})^{-\delta_A}}}}{P_{\operatorname{CP}}+\eta^{-1}P_{\M,l}},
	\end{equation} 
	where $\boldsymbol{P}_\M$ is the vector of transmit powers. Next, we discuss two common formulations for the EE optimization in the SC-MD case: the max-min and sum-EE formulations. 
	
	In the max-min approach, the objective is to maximize the minimum EE of a typical IoT device, and thus the optimization problem is given as
	\begin{equation}
	\label{eq:MaxMin}
	\begin{array}{cl}
	\underset{\boldsymbol{P}_\M}{\text{{maximize}}}~~ \underset{l}{\text{{min}}} 
	&~~ \bar E_{l}(\boldsymbol{P}_\M) \\
	\text{subject to}     
	&~~\sum_l P_{\M,l} \leq \rho \left(\frac{\Delta_\B P_\B}{U_\B}\right) \left(\frac{\lambda_\B}{\lambda_\U}\cdot \frac{\epsilon}{1-\epsilon}\right)^{1/\delta_\G}, \\
	&~~P_\M^{\operatorname{min}}\leq P_{\M,l}\leq P_\M^{\operatorname{max}}~\forall l.\\
	\end{array}
	\end{equation}
	In the sum-EE formulation, the objective is to maximize $\sum_{l=1}^N \bar E_{l}(\boldsymbol{P}_\M)$.
	The EE  expression in (\ref{eq:MaxMin}) is still quasiconcave, and since the objective function is the minimum of quasiconcave functions, it remains quasiconcave. Thus, the max-min problem is quasiconcave. However, for the total EE formulation, the objective function is not necessarily quasiconcave, as the sum of quasiconcave functions does not preserve quasiconcavity. Both approaches can be solved using the generalized Dinkelbach's algorithm \cite{ZapponeJorswieck2015}, which solves the max-min globally, whereas it has become a popular algorithm to solve the sum of ratios, although it does not necessarily arrive at the global optimal solution, if it exists \cite{Schaible2003}.
	
	%------------
	%D. Implementation
	%------------
	\subsection{Implementation and practical considerations}
	Solving the EE problem requires prior knowledge about $\Delta_\B$ and $U_\B$, which are determined by the BS via the transmission mode in LTE or via BS DL precoding in NR. Further, the BS needs to have prior estimates about the network load, i.e., $\lambda_\B$ and $\lambda_\U$. Since IoT devices may have varying PA efficiencies, a nominal value may be used for a given modulation and coding scheme to estimate the total power consumption, i.e., $c(p)$. Estimating the rate, i.e., $r(p)$, is easier as cellular networks already rely on rate-based metrics for link adaptation, e.g., using the channel-equality indicator (CQI) tables, and thus existing methods can be applied. In terms of computational complexity, solving the SC-SD problem has low complexity, and thus the drone can solve it locally, yet the BS must relay the relevant information to the drone. For the multi-drone case, it is more economical to solve the problem at the BS, eliminating the need for the drones to communicate with each other.
	
	Once each IoT cluster receives the nominal transmit power, each IoT device can individually apply fractional power control (FPC) to adapt to the channel quality or path loss. Note that UEs also use FPC, yet the value of $P_\U$ does not affect the performance of IoT devices or the EE optimization problem since IoT devices do not interfere with UL UEs in the proposed protocol.
	
	Finally, there remain practical challenges related to deploying drones and synchronizing UEs and IoT devices. For example, the authors in \cite{Motlagh2017} have already developed an IoT platform that uses drones for crowd surveillance, yet scaling such a platform requires innovative solutions to enable extended hours of operations and autonomous control of a large number of drones. Furthermore, similar to a BS synchronizing multiple UEs over the same time-frequency slot for MIMO spatial-multiplexing, the BS must extend its capability to synchronize UEs and IoT devices using UL scheduling grants in order to implement the proposed TDD protocol.

	%---------------------------------------------------------------------------------
	%                         VI. Simulation Results
	%---------------------------------------------------------------------------------
	\section{Simulation Results}\label{sec:simulations} 
	Unless otherwise stated, we use the simulation parameters given in Table \ref{tab:parameters}. In each spatial realization of the network, we generate BSs, UEs, and IoT devices according to their distributions. Each drone then moves to their predetermined stop points. We then generate the channel power gains according to their distributions and compute the large-scale fading for all links to compute their SINR, where a thermal noise power, with spectral density $-174$dBm/Hz, is considered at all receivers. Then the SE of UEs is computed via (\ref{eq:SE}) and the EE IoT devices is computed via (\ref{eq:EE}) for the different transmission protocols. We note that we use the actual load and the 3GPP LOS probability, instead of their mean, to compute the performance metrics. 
	
	%------------
	%A. Validation of the stochastic optimization frameworks
	%------------
	\subsection{Validation of the theoretical analysis} 
	We first validate the theoretical analysis and the stochastic optimization problem, focusing on SC-SD and SC-MD scenarios. We compare the EE performance under the proposed nominal transmit power with that achieved under a coverage-maximizing scheme, where the IoT device transmits at a maximum power of $P_\M=23$dBm \cite{3GPP2015a}.
	
	%-------
	%1. Impact of IoT transmit power
	%-------
	\subsubsection{Impact of IoT transmit power on the EE} 
\begin{figure}[t!]
		\centering
		\begin{subfigure}{.45\textwidth}
			\center
			\includegraphics[width=3.0in]{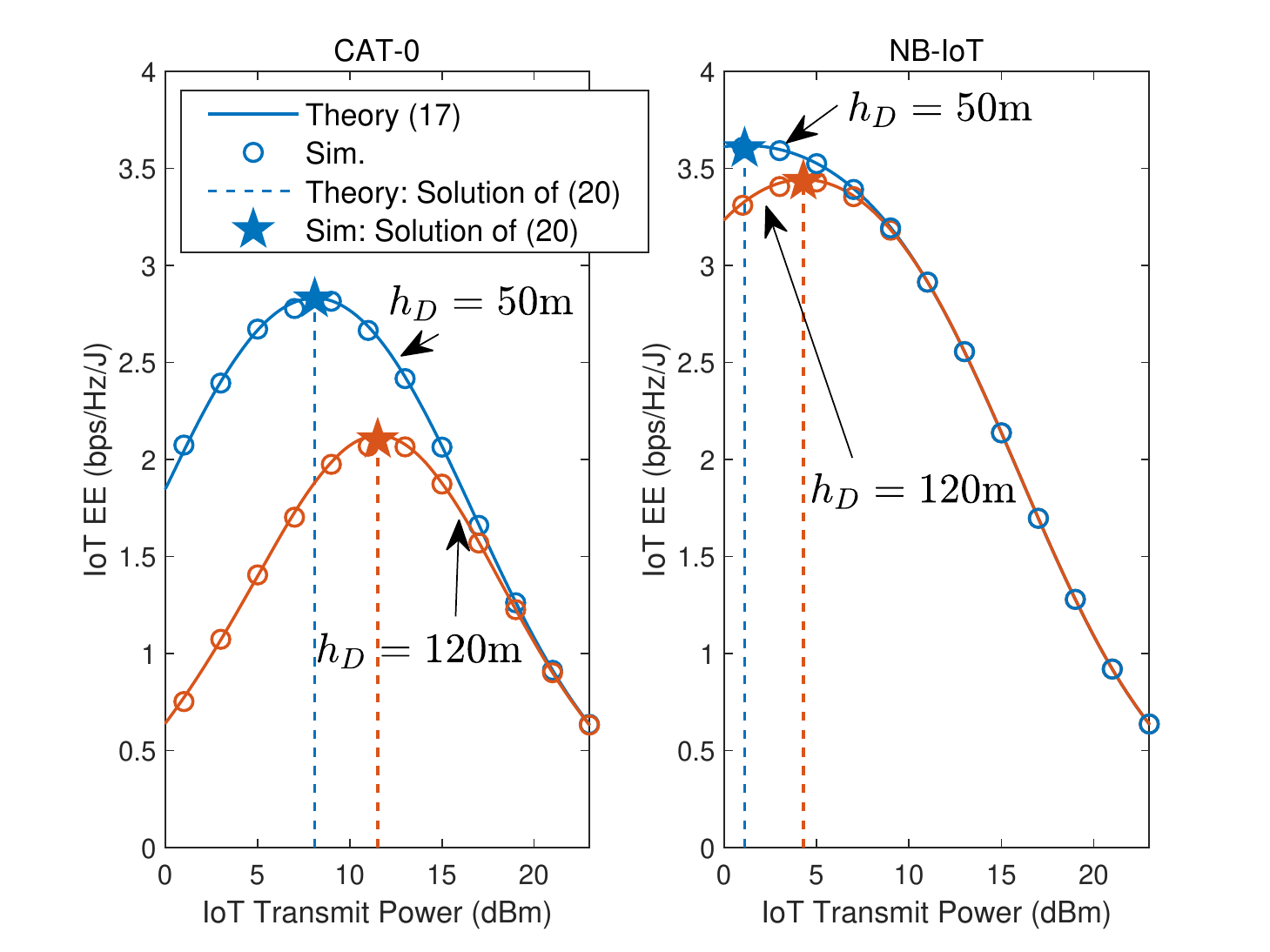}
			\caption{Energy-efficiency performance}
			\label{fig:EE_vs_Pm}
		\end{subfigure}	~
		\begin{subfigure}{.45\textwidth}
			\center
			\includegraphics[width=3.0in]{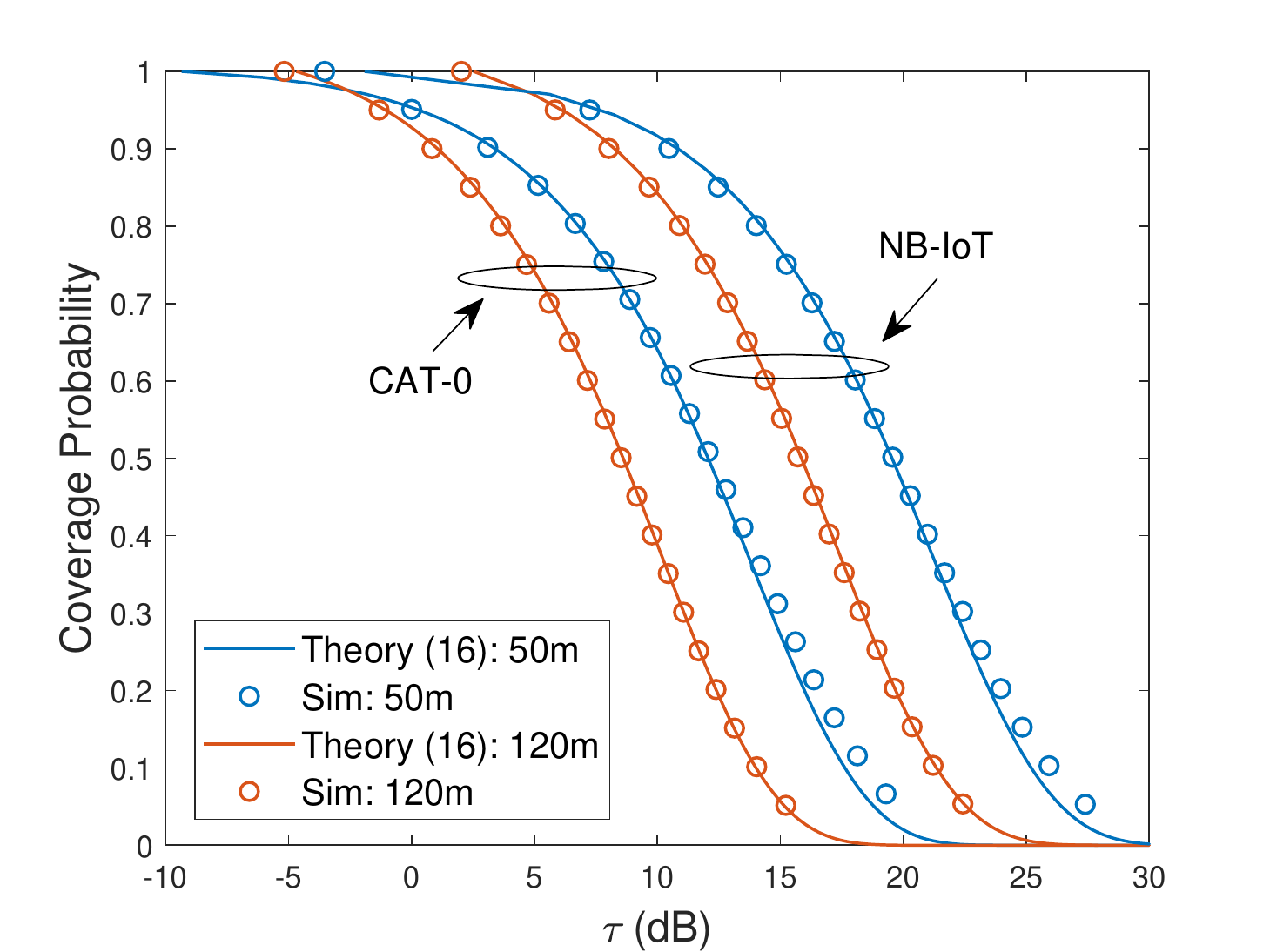}
			\caption{Coverage probability}
			\label{fig:Cov_sim_vs_theo}
		\end{subfigure}	\caption{Validation of the theoretical EE and coverage.}
		\label{fig:EEvsPm}
		\vspace{-.15in}
	\end{figure}

	We consider two IoT categories: CAT-0 and NB-IoT. In the former, the IoT device shares the entire band with the UE, i.e., $W=20$MHz and $P_\B=46$dBm. In the latter, the IoT only shares a single resource block, i.e., $W=180$KHz and $P_\B=32$dBm. For interference protection, we assume $\epsilon=0.5$ and $\rho_{\operatorname{dB}}=-6$dB, i.e., the median ISR should not exceed $-6$dB. Note that this ISR threshold is commonly used as a protection criterion in FCC's regulations \cite{FCC2017}, and it corresponds to an SINR degradation of 1dB.
	The performance of EE under different transmit powers is shown in Fig. \ref{fig:EE_vs_Pm}.  It is observed that the theoretical expression in (\ref{eq:IoTEE}) matches well with Monte Carlo simulations. Further, the EE is significantly improved using the proposed SC-SD framework compared to the max-power scheme, e.g., the EE in CAT-0 of the proposed SC-SD framework is 4.5x and 3.3x that of max-power for $h_\D=50$m and $h_\D=120$m, respectively. This follows because the drone's location is optimized to minimize the 2D distance to the IoT device, requiring low transmit power for reliable coverage. Third, increasing the drone's height has two effects: an increase in the optimal transmit power and a decrease in the EE. These follow because as $h_\D$ increases, the received signal power decreases more rapidly than interference, degrading the coverage. Thus, the IoT device must transmit at higher power to combat the degradation, decreasing its EE.  Finally, the NB-IoT operation is more efficient than CAT-0 since in the former the IoT device shares a smaller number of carriers with the UE, reducing the interference. In Fig. \ref{fig:Cov_sim_vs_theo}, we validate the coverage approximation in (\ref{eq:IoTCov}), which is shown to be in good agreement with Monte Carlo simulations. It is evident that using drones at lower altitudes is critical to provide high coverage, making it an alternative to IoT devices transmitting at a higher UL power.
	
	%-------
	%2. Impact of UAV-BS coordination
	%-------
	\subsubsection{Validation of BS transmit power optimization and the SC-MD case}
	We validate the generalizations of the SC-SD. Here, we consider CAT-0 parameters  (similar trends are observed for NB-IoT and hence omitted).
	
\begin{figure}[t!]
		\centering
		\begin{subfigure}{.48\textwidth}
			\center
			\includegraphics[width=3.25in]{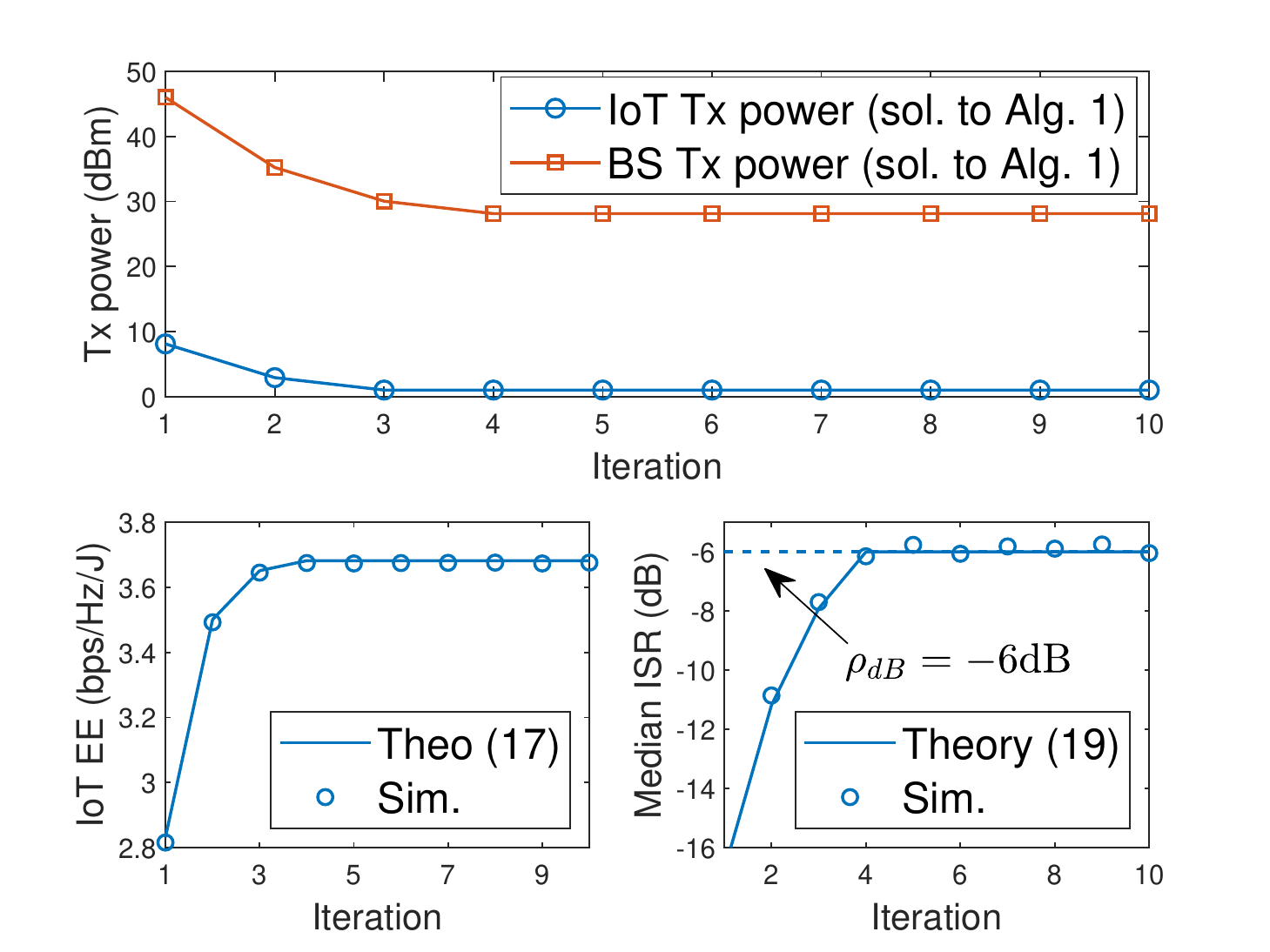}
			\caption{Impact of optimizing BS power ($\lambda_{\M}=50$m)}
			\label{fig:BSopt}
		\end{subfigure}~
		\begin{subfigure}{.48\textwidth}
			\center
			\includegraphics[width=3.25in]{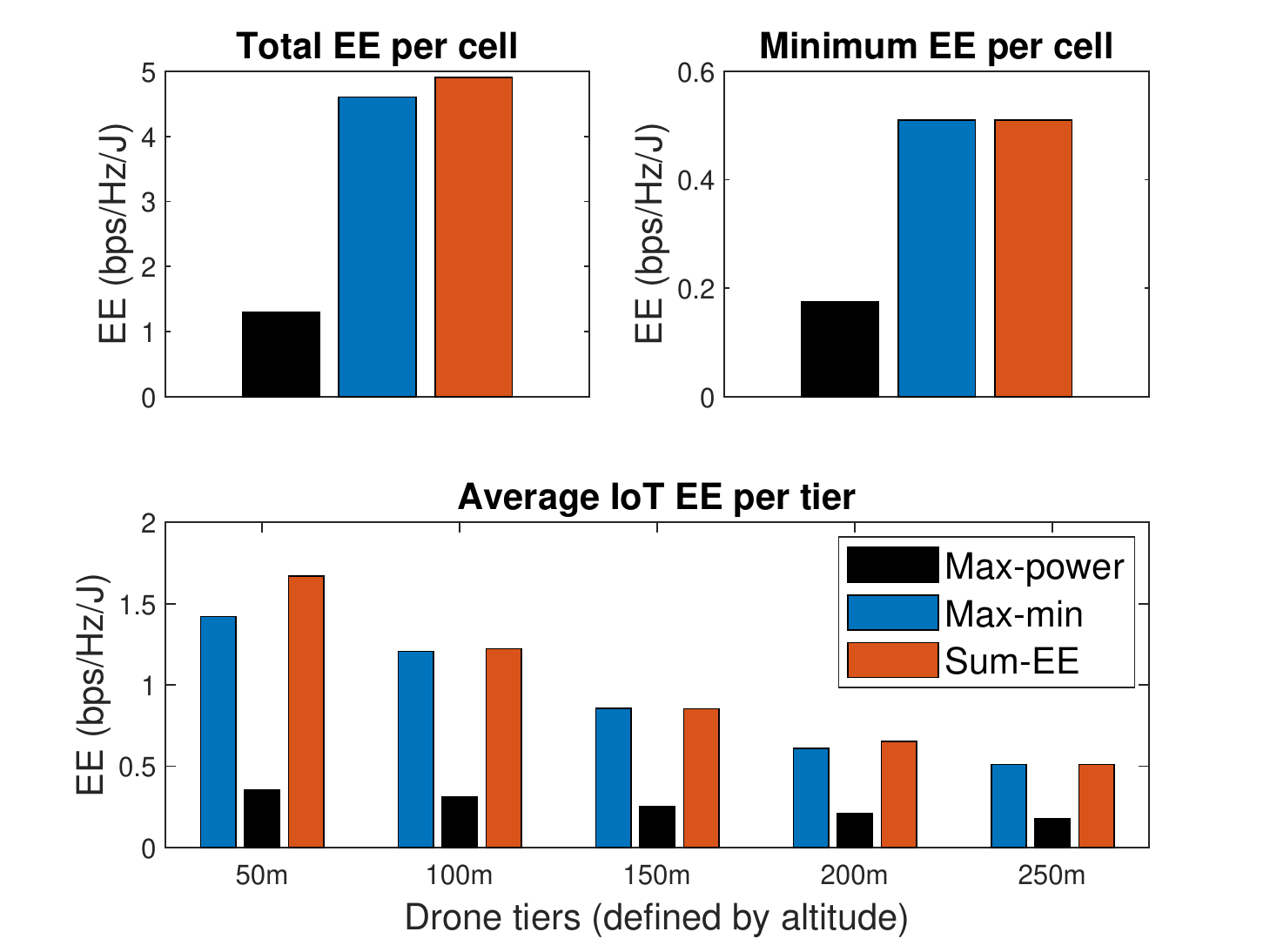}
			\caption{EE performance in the SC-MD case}
			\label{fig:EE_SCMD}
		\end{subfigure}
		\caption{Validation of generalizations to the SC-SD case (CAT-0 parameters and $\rho_{\operatorname{dB}}=-6$dB).}
		\label{fig:generalSCSD}
		\vspace{-.15in}
	\end{figure}

	Fig. \ref{fig:BSopt} illustrates the impact of optimizing the BS transmit power using Alg. \ref{alg:BSopt}. It is shown that the BS's transmit power decreases at first since the ISR constraint is satisfied with strict inequality. This allows the IoT device to further decrease its transmit power, which in return improves its EE, until the constraint is satisfied with equality. For instance, the EE improves approximately by $30\%$ compared to the case without BS power optimization. We note that optimizing $P_\B$ may not be applicable in cases where stricter UE protections are required.
	
	Next, we study the EE performance in the presence of multiple drones in the same cell, where we assume they can fly at one of the these altitudes: $\boldsymbol{h}_\D=[50,100,150,200,250]$m. Fig. \ref{fig:EE_SCMD} shows the performance of \emph{Max-min} and \emph{Sum-EE} schemes, where power allocation is done using the Generalized Dinkelbach’s algorithm. It is evident that the EE is improved compared to max-power. We further show the EE of IoT devices that belong to the different drones. It is observed that the \emph{Max-min} solution aims to improve the EE of devices connected to the drone with the highest altitude as this tier achieves the lowest EE. In contrast, the \emph{Sum-EE} formulation favors the drones with lower altitudes as they provide higher EE for IoT devices. The disparity between both formulations at the lowest and highest altitudes can increase with stricter ISR threshold.
	
	%------------
	%B. Performance comparison with existing protocols
	%------------
	\subsection{Performance comparison with existing protocols}
	In this section, we compare the performance of the proposed protocol with other ones in large networks, i.e., many cells, and hence the results are obtained via Monte Carlo simulations. To evaluate the impact of IoT coexistence on the UEs' spectral efficiency, we consider a benchmark scheme without IoT devices. The proposed protocol is then compared to the following coexistence schemes: (i) a standard spectrum sharing protocol, (ii) orthogonal-based protocol that uses frequency partitioning, and (iii) the proposed protocol, but with terrestrial aggregators that are deployed at the clusters' centroid and IoT devices transmitting at maximum power. Further, we also study the performance of the proposed protocol with the optimal $P_\M$ that maximizes the EE over the entire network, i.e., all cells and drones instead of just the SC-SD case. We obtain $P_\M$ using extensive exhaustive simulations. We consider CAT-0 IoT devices (similar trends are observed for NB-IoT), $\lambda_\U=50\lambda_\B$ and $\lambda_\Cl=\lambda_\D=5\lambda_\B$. Unless otherwise stated, we assume $\delta=1$, $W_\U=0.5W$, and $h_\D=50$m. 
	
	%-------
	%1. UE performance comparison
	%-------
	\subsubsection{UE performance comparison} 
\begin{figure}[t!]
		\centering
		\begin{subfigure}{.45\textwidth}
			\center
			\includegraphics[width=3.0in]{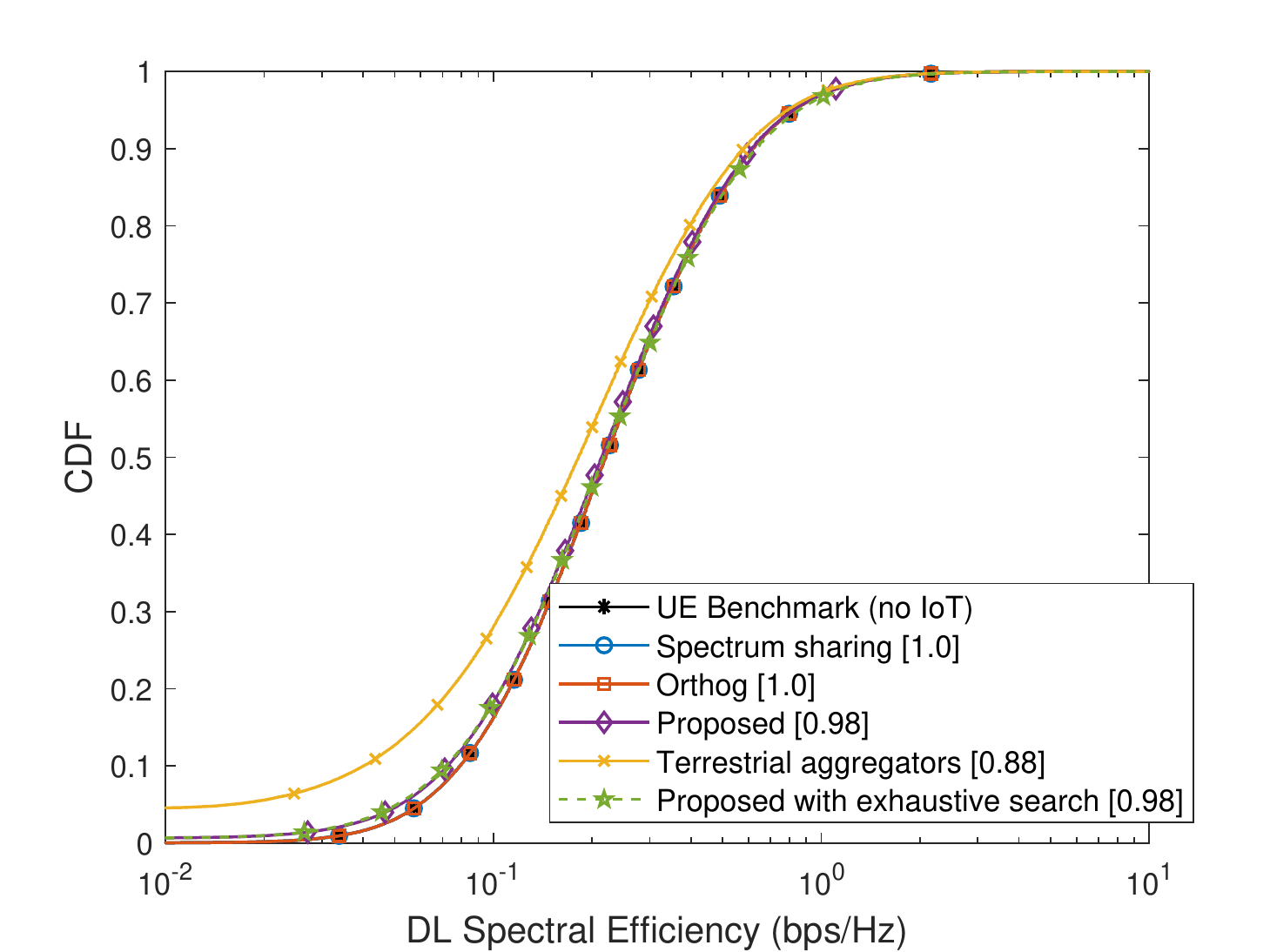}
			\caption{Distribution of DL SE}
			\label{fig:SEdlCDF}
		\end{subfigure} ~~ 
		\begin{subfigure}{.45\textwidth}
			\center
			\includegraphics[width=3.0in]{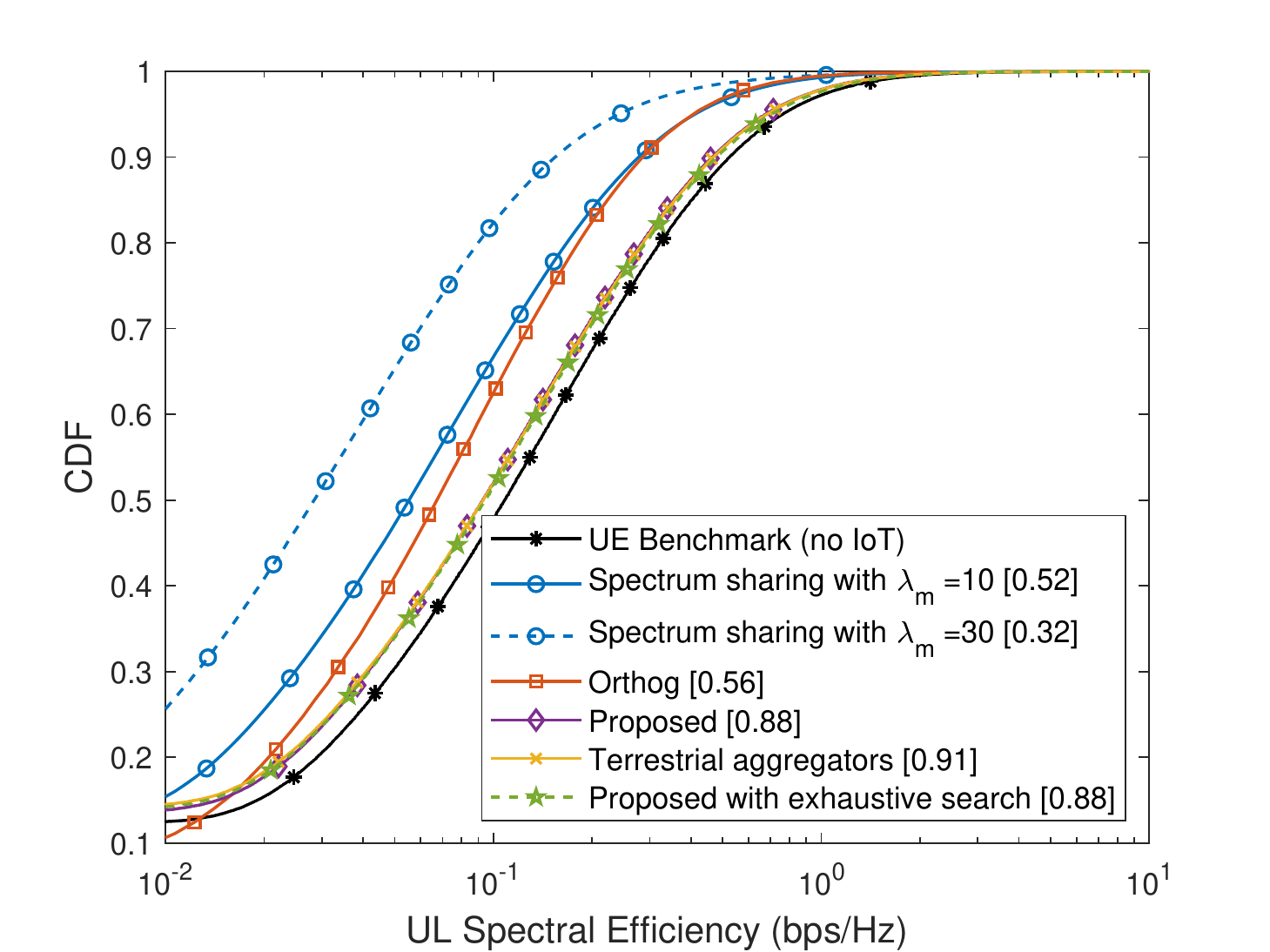}
			\caption{Distribution of UL SE}
			\label{fig:SEulCDF}
		\end{subfigure}
		\caption{UE spectral efficiency under different protocols (CAT-0 parameters and $\rho_{\operatorname{dB}}=-6$dB).}
		\label{fig:UESE}
		\vspace{-0.15in}
	\end{figure}

	We first study the performance of the UE in the DL and the UL under different protocols.  We note that both the orthogonal-based and the aggregator-based protocols decouple the UE UL performance from the IoT density as in the former IoT devices use different frequency blocks, and in the latter IoT devices connect to aggregators instead of BSs. This is not the case for the sharing-based protocol, where we study the UE's performance for $\lambda_\M=10$ and $\lambda_\M=30$. Fig. \ref{fig:SEdlCDF} and Fig. \ref{fig:SEulCDF} show the distribution of the DL SE and UL SE across all UEs, respectively. We also show in the legend the mean SE relative to the benchmark, i.e., the ratio of the mean SE under the given protocol to the mean SE in the absence of IoT devices. Since our model considers IoT devices to only operate in the UL, the DL SE of spectrum sharing and frequency partitioning protocols is the same as the benchmark. For the proposed protocol, it is shown that the DL degradation is minimal since UAVs are used and the transmit power is optimized to limit the IoT interference. More importantly, the proposed protocol outperforms spectrum sharing and frequency partitioning in the UL, e.g., the relative mean UL SE of the proposed protocol is improved by 2.7x compared to spectrum sharing ($\lambda_\M=30$). Finally, using terrestrial aggregators improves the UL SE, similar to using drones, yet the DL SE is still degraded by 10\% compared to the proposed protocol.
	
	%-------
	%2. IoT performance comparison
	%-------
	\subsubsection{IoT performance comparison} 
\begin{figure}[t!]
		\centering
		\begin{subfigure}{.45\textwidth}
			\center
			\includegraphics[width=3.0in]{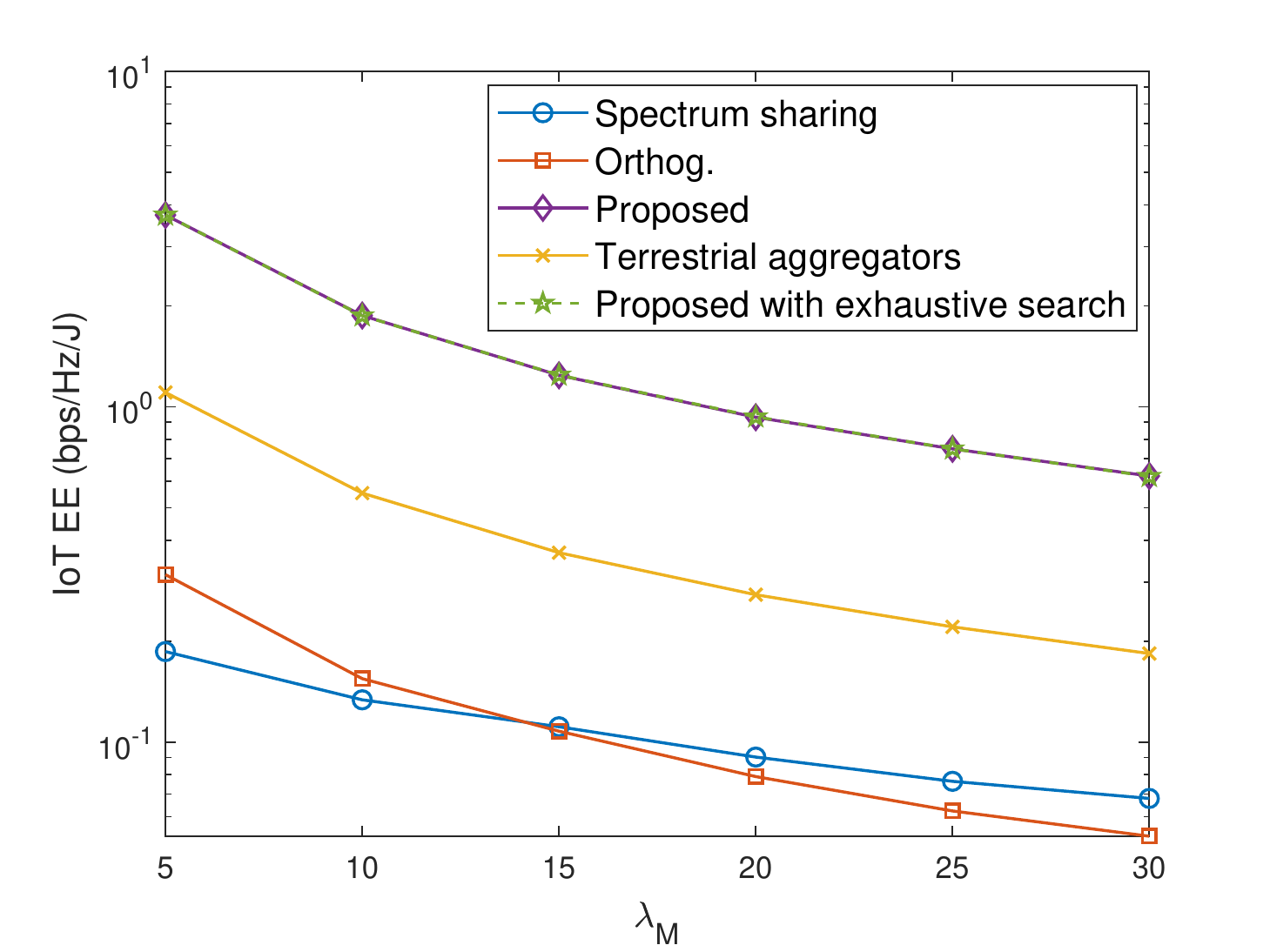}
			\caption{Variations of IoT density ($h_\D=50$m)}
			\label{fig:EEvsNoIoT}
		\end{subfigure} ~
		\begin{subfigure}{.45\textwidth}
			\center
			\includegraphics[width=3.0in]{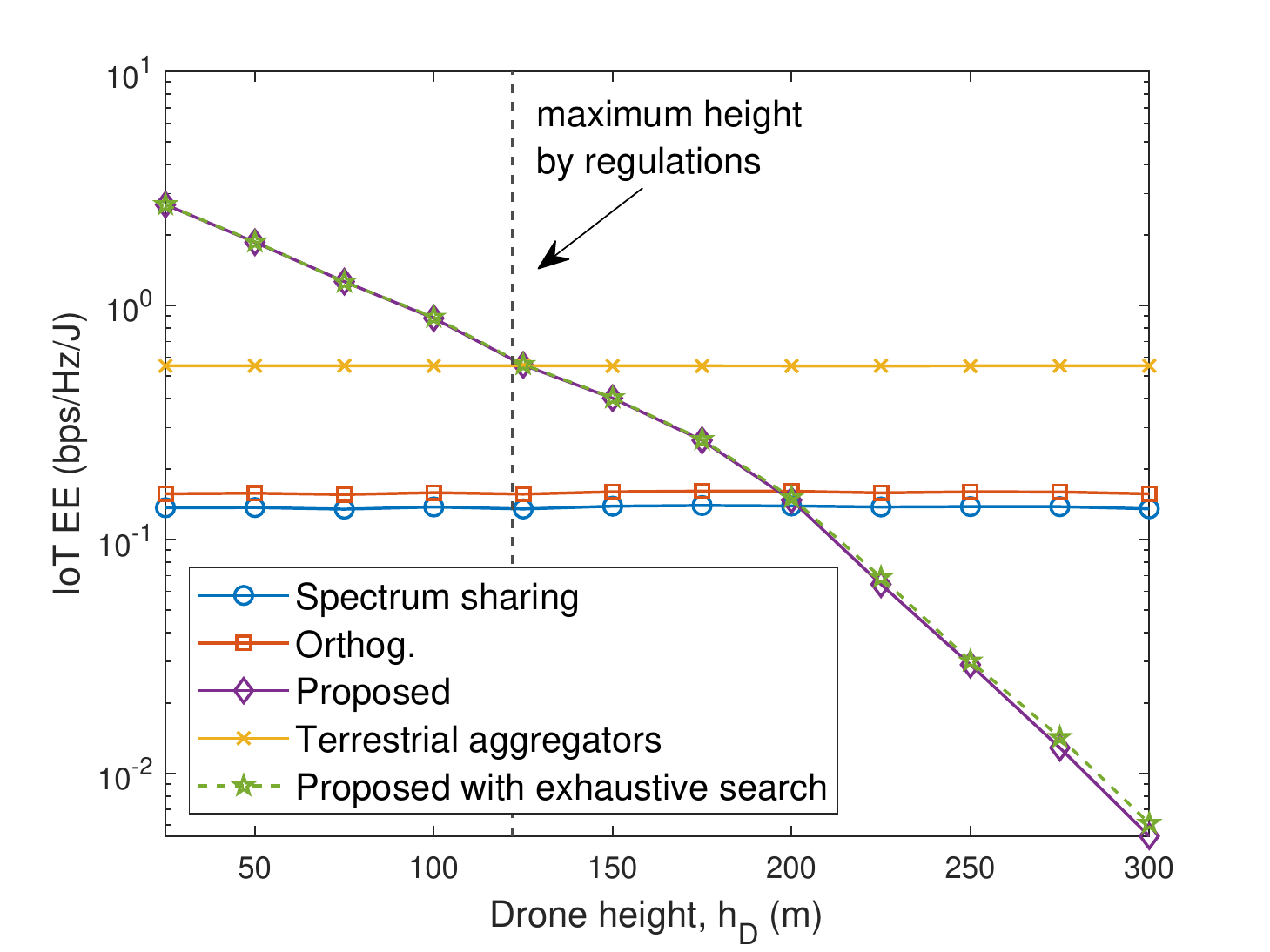}
			\caption{Variations of drone's height ($\lambda_\M=10$)}
			\label{fig:EEvsHeight}
		\end{subfigure}
		\caption{IoT EE performance under different protocols.}
		\label{fig:IoTEE}
		\vspace{-0.25in}
	\end{figure}

We then study the EE of the IoT device under the different protocols. In Fig. \ref{fig:EEvsNoIoT}, the EE is shown for different densities of IoT devices. As expected, as the number of IoT devices increases, the EE decreases under all protocols. Yet, the proposed protocol achieves the highest EE and scales better with $\lambda_\M$ compared to existing ones. It is also observed that it is beneficial to use UAVs over terrestrial aggregators as the former provides higher LOS, i.e., the EE improves by 3x when aerial aggregators are used instead of terrestrial ones. In Fig. \ref{fig:EEvsHeight}, we study the EE for different drone's altitudes. We show the performance under the existing protocols and the proposed one with terrestrial aggregators, which do not depend on $h_\D$, for reference. It is observed that the proposed protocol is beneficial for lower altitudes. At very high altitudes, the received signal power is lower, and the LOS probability with interfering BSs, from different cells, increases. We note that in many regions, e.g., North America, Europe, China, etc., the maximum legal altitude for drones is roughly 120m (400ft), and thus the proposed solution is superior in practical scenarios. 
	
\begin{figure}[t!]
		\centering
		\begin{subfigure}{.45\textwidth}
			\center
			\includegraphics[width=3.0in]{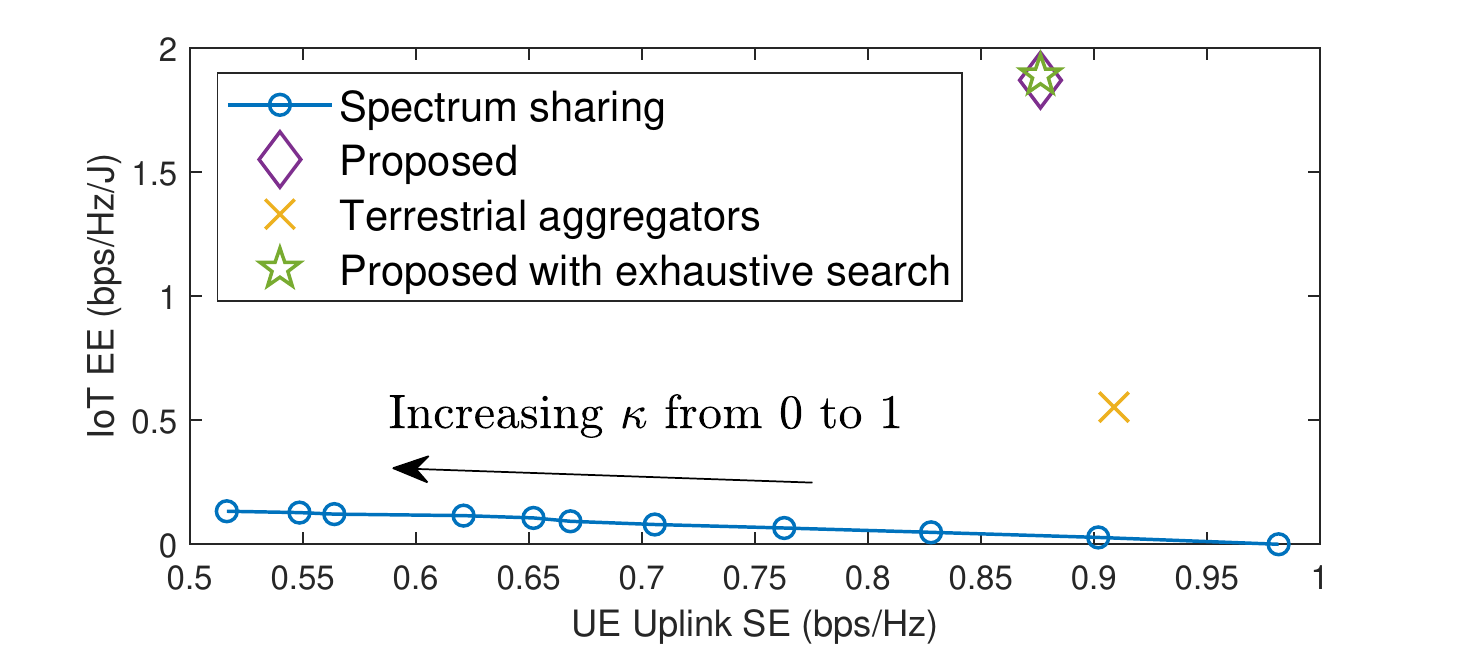}
			\caption{Varisations of ACB threshold}
			\label{fig:ACB}
		\end{subfigure}~
		\begin{subfigure}{.45\textwidth}
			\center
			\includegraphics[width=3.0in]{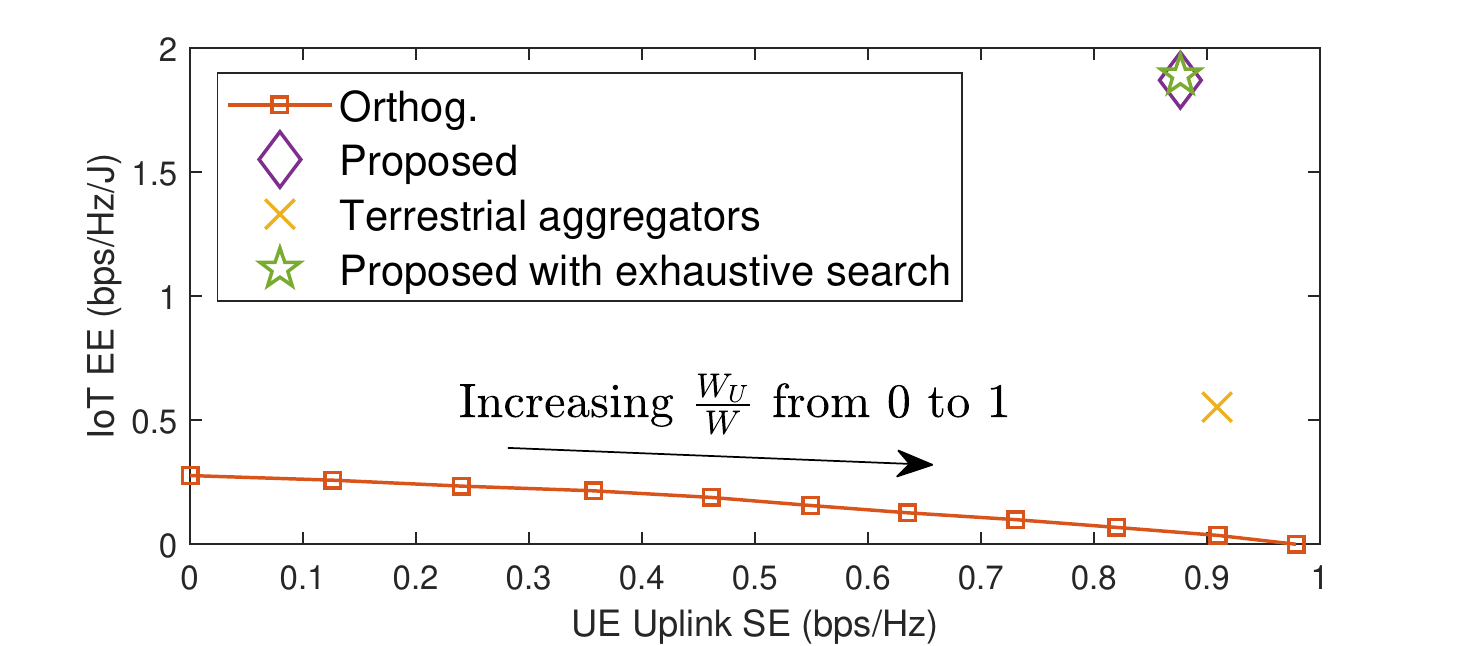}
			\caption{Variations of frequency partitioning ratio}
			\label{fig:resourceSplitting}
		\end{subfigure}
		\caption{IoT energy-efficiency vs. UE spectral efficiency ($\lambda_\M=10$ and $h_\D=50$m).}
		\label{fig:existingComparison}
		\vspace{-.25in}
	\end{figure}

	%-------
	%3. Impact of ACB and resource splitting
	%-------
	%\subsubsection{Impact of ACB and resource splitting}
	Fig. \ref{fig:existingComparison} shows the IoT EE versus the UE SE in the UL under different ACB thresholds $\kappa$ (Fig. \ref{fig:ACB}) and different frequency allocation ratios $W_{\U}$ (Fig. \ref{fig:resourceSplitting}). The performance of the proposed protocol, which does not depend on these parameters, is also shown for reference. It can be seen that existing protocols have operating points with higher UE SE performance in comparison with proposed ones. However, high ACB threshold and frequency partitioning ratio are needed, and thus the IoT device will be limited with time and frequency resources, respectively. Finally, it is shown that optimizing $P_\M$ over the SC-SD case leads to a nearly identical performance to exhaustive search, yet the latter requires extensive simulations to solve.
	
	\subsection{IoT device lifetime comparison}
	In this section, we follow the 3GPP evaluation methodology to compute the lifetime of IoT devices for the different schemes \cite{3GPP2015e,Hattab2018}. In particular, the IoT device sends $N_{\operatorname{rep}}$ reports per day to the network. For each report, the device operates in the following stages: standby, idle, transmission, and reception. Let $P_{\operatorname{S}}$, $P_{\operatorname{I}}$, and $P_{\operatorname{RX}}$ denote the power consumption of the standby, idle, and reception stages. Similarly, let $T_{\operatorname{S}}$, $T_{\operatorname{I}}$, and $T_{\operatorname{RX}}$ be their durations. For a fair comparison, we assume the energies consumed for these three stages are the same across the schemes, i.e., the schemes only differ in the energy consumed during the transmission stage as our focus is on the UL. The transmission duration is
	\begin{equation}
	T_{\operatorname{TX}}^{(\nu)} = \frac{B/ \beta_{\M}^{(\nu)}}{\bar\sum_{k=1}^{K_\M} \log_2(1+\tau_{\M,k}) \boldsymbol{1}(\tau_{\M,k+1}\geq \gamma_\M^{(\nu)}\geq \tau_{\M,k})},
	\end{equation}
	where $B$ is the total size of the data transmitted per report, which includes the connection request, the data packet, and the acknowledgment \cite[Table 4]{3GPP2015e}. The superscript $\nu$ denotes the scheme used. Similarly, $P_{\operatorname{TX}}^{(\nu)}=P_{\operatorname{CP}}+\eta^{-1}P_{\M}^{(\nu)}$. Thus, the total energy consumed per day is given as \cite{3GPP2015e}
	\begin{equation}
	E_{\operatorname{IoT}}^{(\nu)} = N_{\operatorname{rep}} \left(T_{\operatorname{TX}}^{(\nu)} P_{\operatorname{TX}}^{(\nu)} + T_{\operatorname{RX}}P_{\operatorname{RX}} + T_{\operatorname{I}}P_{\operatorname{I}}\right) + T_{\operatorname{S}}P_{\operatorname{S}}.
	\end{equation}
	Let the IoT battery's capacity be $C_{\operatorname{IoT}}$ Wh, then the device lifetime in years is given as $Y^{(\nu)} = \frac{C_{\operatorname{IoT}}}{E_{\operatorname{IoT}}^{(\nu)}}\times\frac{3600}{365}$. Note that increasing the IoT transmit power is expected to improve the SINR, increasing the rate and decreasing $T_{\operatorname{TX}}^{(\nu)}$. Yet, this comes at the expense of increased power consumption during the transmission stage, i.e., higher $P_{\operatorname{TX}}^{(\nu)}$. 
	
	For the simulation set-up, we use a more realistic deployment and channel model. Specifically, we consider a hexagonal deployment of BSs, and consider the NR 3D-UMa channel model for ground-to-ground links \cite{3GPP2017d} and the UMa-AV model for ground-to-air links \cite{3GPP2017a}. These models assume multi-slope path loss with different attenuation, depending on whether the link is LOS or non-LOS, as well as consider log-normal shadowing. We assume the antenna height of the BS and IoT devices are 30m and 1.5m, respectively, and the operating channel is centered at 2GHz. We use the battery lifetime parameters given in Table \ref{tab:lifetime} \cite{3GPP2015e}. Finally, we allow each IoT device to apply fractional power control on top of the optimized nominal transmit power to compensate for path loss, where we consider a factor of $0.3$.
	
\begin{figure}[t!]
		\centering
		\scriptsize
		\includegraphics[width=3.0in]{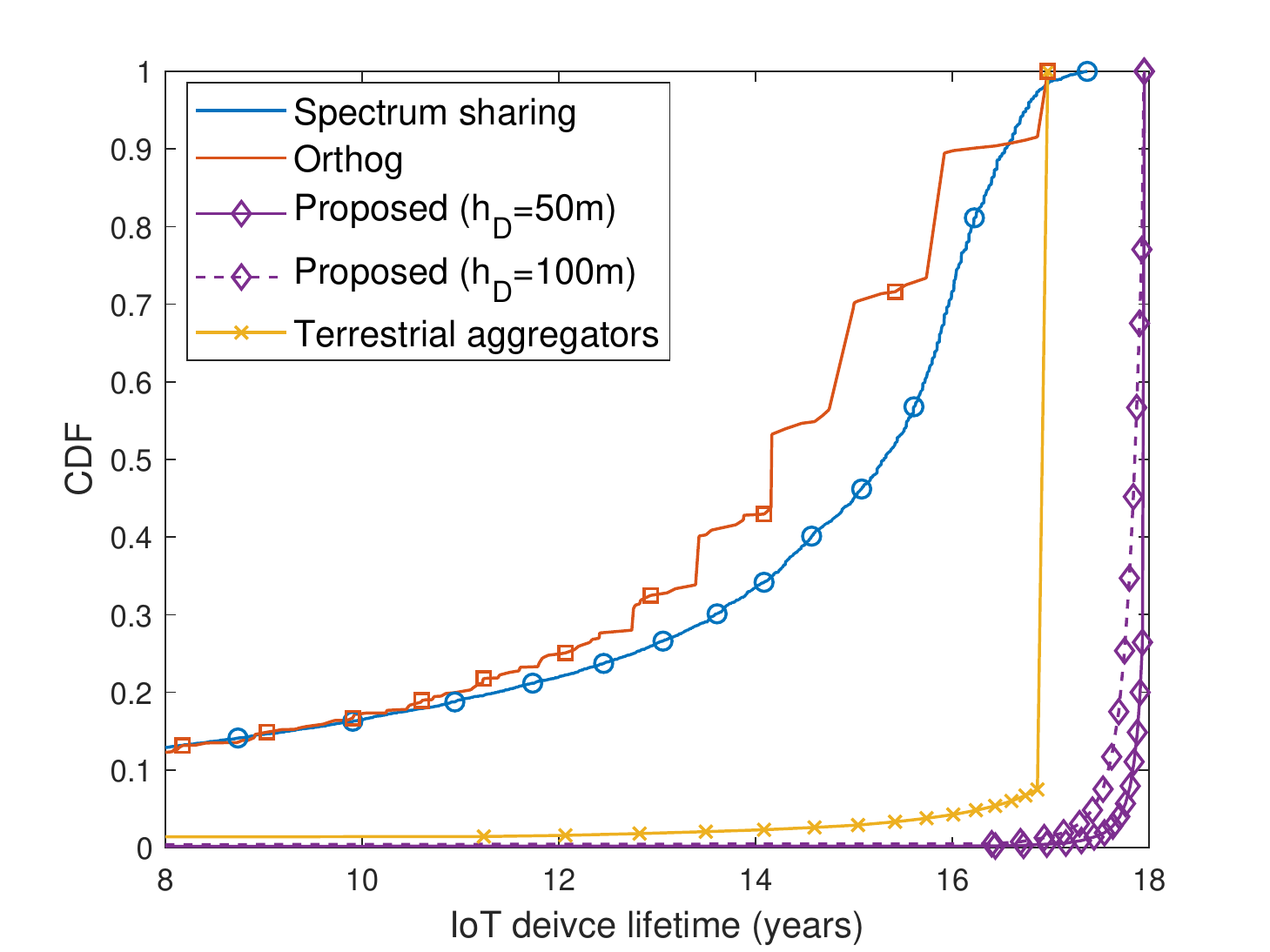}
		\caption{Distribution of IoT device lifetime.}
		\label{fig:lifetimeCDF}
		\vspace{-.25in}
	\end{figure}

	Fig. \ref{fig:lifetimeCDF} shows the CDF of the IoT device lifetime for each scheme. We have the following observations. First, the improvements in EE under the proposed scheme are translated into tangible enhancements to the IoT device lifetime, e.g., the median lifetime is increased by more than three years compared to spectrum sharing and frequency partitioning schemes. Second, due to the aggregators' proximity to IoT devices, whether aerial or terrestrial, the variance in lifetime across devices is lower under compared to schemes without aggregators. When UAVs are used instead of terrestrial aggregators, the variance is further reduced thanks to the better LOS conditions. Finally, as the drone altitude increases, the IoT device lifetime slightly decreases, i.e., the performance gain of using drones over terrestrial aggregators is reduced.
	
	\begin{table}[!t]
		\caption{Battery lifetime parameters \cite{3GPP2015a,3GPP2015e}}
		\label{tab:lifetime}
		\centering
		\scriptsize
		\begin{tabular}{|c|l|}
			\hline
			\textbf{Description }  				&  \textbf{Parameters}\\\hline
			\multirow{1}{*}{IoT device} & NB-IoT with $W=180$KHz,  $C_{\operatorname{IoT}}=5$Wh, 		$B=229$bytes \cite[Table 4]{3GPP2015e}, and $N_{\operatorname{rep}}=12$\\\hline
			\multirow{1}{*}{Powers}     & $P_{\operatorname{RX}}=90$mW, $P_{\operatorname{I}}=3$mW, 
			and $P_{\operatorname{S}}=0.015$mW \cite[Table 1]{3GPP2015e}\\\hline
			\multirow{1}{*}{Durations} 	& $T_{\operatorname{RX}}=565$ms, 
			$T_{\operatorname{I}}=22451$ms, and $T_{\operatorname{S}}=86400$s 	\cite[Table 6]{3GPP2015e}		\\	 	\hline	
		\end{tabular}
		\vspace{-0.25in}
	\end{table}
	
	%---------------------------------------------------------------------------------
	%                         VI. Conclusions
	%---------------------------------------------------------------------------------
	\section{Conclusions}\label{sec:conclusion}
	In this work, we have proposed a TDD protocol for a shared spectrum access between massive IoT and cellular UEs with UAVs acting as data aggregators. Using stochastic geometry, it is shown that the protocol improves the average allocated resources of IoT devices and UEs compared to resource splitting and ACB, yet UEs experience additional interference from IoT devices. Thus, we have optimized the nominal transmit power of the IoT device to maximize its energy-efficiency while constraining the interference on UEs. The optimal nominal power can be then biased by each IoT device individually using uplink power control when additional path loss compensation is needed. Simulation results show that the proposed protocol significantly improves the EE, and hence the IoT device lifetime, thanks to the drones proximity to the IoT device, while still protecting the UEs thanks to the optimized IoT transmit power.

	The key insights gleaned from this work are as follows. First, it is beneficial for drones to fly at lower altitudes as higher altitudes increase the LOS with interferers, forcing the IoT device to increase its transmit power and degrading its EE. In case drones, belonging to the same BS, fly at different altitudes, then maximizing the minimum EE and the total EE amount to prioritizing devices connected to drones at the highest and the lowest altitudes, respectively. Second, the extended coverage mode, where IoT devices transmit at maximum power, is not necessarily energy efficient  as the gain in coverage does not outweigh the loss in power consumption.

	%---------------------------------------------------------------------------------
	%                         Appendices
	%---------------------------------------------------------------------------------
	\appendix
	
	\subsection{Proof of Theorem \ref{theo:UEcoverage}}\label{app:theorem1}
	Recall that the aggregate interference at the UE is given as 
	\begin{equation}
	I_\U= \underset{I_{\U,\B}}{\underbrace{\sum_{y_b\in\Phi_\B'} f_b y_b^{-\alpha_\G}}}+\underset{I_{\U,\M}}{\underbrace{\sum_{z_m\in\Phi_\M'} \frac{P_\M}{P_\B/U_\B} f_m  z_m^{-\alpha_\G}}},
	\end{equation}
	where we have normalized the interference by $L_0P_\B/U_\B$. Using the Gil-Pelaez Inversion theorem \cite{GilPelaez1951} to evaluate the interference CDF, we get $
	F_{I_{\U}}\left(\frac{g_\B}{\tau x^{\alpha_\G}}\right) = \frac{1}{2}-\frac{1}{\pi} \int_{0}^{\infty}\operatorname{Im}\left\{\varphi_{I_\U}(x^{\alpha_\G}t)e^{-jt g_\B/\tau}\right\}dt$, 
	where $\varphi_{I_\U}(x^{\alpha_\G}t)= \mathbb{E}_{I_\U}\left[\exp\left(jtx^{\alpha_\G} I_\U\right)\right]$ is the characteristic function (CF), which is given as
	\begin{equation}
	\begin{aligned}
	\varphi_{I_\U}(x^{\alpha_\G}t) &\stackrel{(a)}{=} \textstyle \mathbb{E}_{I_{\U,\B}}\left[e^{\left(jtx^{\alpha_\G} \sum_{y_b\in\Phi_\B'}f_b y_b^{-\alpha_\G}\right)}\right]\times
	\textstyle  \mathbb{E}_{I_{\U,\M}}\left[e^{\left(jtx^{\alpha_\G} \sum_{z_m\in\Phi_\M'}\frac{P_\M}{P_\B/U_\B}f_m  z_m^{-\alpha_\G}\right)}\right], 
	\end{aligned}
	\end{equation}
	where $(a)$ follows since interfering BSs are independent from the interfering IoT devices. Using the probability-generating functional of the HPPP process \cite{Haenggi2012}, it can be shown that $ \textstyle \varphi_{I_{\U,\B}}(x^{\alpha_\G}t) = \exp\left(\pi\lambda_\B x^2 (1-\mathbb{E}_{f_b}[\Omega(f_b,\delta_\G,t)])\right)$,
	where $\Omega(f_b,\delta_\G,t)={}_1F_1(-\delta_\G;1-\delta_\G;j t f_b)$, and $_1F_1(a;b;c)$ is the confluent hypergeometric function \cite{JeffreyZwillinger2014}.  Similarly, we have
	\begin{equation}
	\begin{aligned}
	\textstyle \varphi_{I_{\U,\M}}(x^{\alpha_\G}t) 
	&\stackrel{(a)}{=} e^{-2\pi\lambda_\D \int_{0}^{\infty}y \left(1-\mathbb{E}_{f_m}\left[e^{jt \frac{U_\B P_\M f_m x^{\alpha_\G}}{P_\B y^{\alpha_\G}} }\right]\right)dy}\\
	%	&\stackrel{(b)}{=} e^{-\delta_\G\pi\lambda_\D x^2 \int_{0}^{\infty}l^{-(1+\delta_\G)} \left(1-\frac{1}{1-j t l \frac{P_\M}{P_\B/U_\B}  }\right)dl}\\
	&\stackrel{(b)}{=}\exp\left(-\pi\lambda_\D x^2 \hat P_{\B,\M}^{\delta_\G}(-jt)^{\delta_\G}\right) ,
	\end{aligned}
	\end{equation}
	where $(a)$ follows using the fact that the set of interfering IoT devices can be modeled as a set of Poisson interferers with density of $\lambda_\D$ as one IoT device is scheduled per drone per time-frequency slot, and $(b)$ follows using the CF of $f_m$ and then using the substitution $l=x^{\alpha_\G}y^{-\alpha_\G}$. Here, the integral lower limit is zero as the set of interfering IoT devices is independent of the UE's location, i.e., no protection zone is present unlike the case of interfering BSs which cannot be closer than the tagged BS. 
	To summarize, we have $
	\varphi_{I_{\U}}(x^{\alpha_\G}t) = \exp\left(\pi \lambda_\B x^2 \left(1-\mathbb{E}_f[\Omega(f_b,\delta_\G,t)]-(\lambda_\D/\lambda_\B)\hat P_{\B,\M}^{\delta_\G}(-jt)^{\delta_\G}\right)\right)$. Thus, the coverage becomes
	\begin{equation}
	\label{eq:cov}
	\begin{aligned}
	\mathbb{C}^{\operatorname{P}}_{\U,\operatorname{DL}}(\tau)
	&= \frac{1}{2}-2\lambda_\B \int_{0}^{\infty} \frac{1}{t} \operatorname{Im} \left\{\varphi_g(-t/\tau)\Xi(t)\right\}dt,
	\end{aligned}
	\end{equation}
	where $\varphi_{g_\B}(-t/\tau) =  \frac{1}{\left(1+jt/\tau\right)^{\Delta_\B}}$ is the CF of $g_\B$ and $\Xi(t) =\frac{(2\pi\lambda_\B)^{-1}}{\mathbb{E}_f[\Omega(f_b,\delta_\G,t)]+(\lambda_\D/\lambda_\B)\hat P_{\B,\M}^{\delta_\G}(-jt)^{\delta_\G}}$. 
	%\begin{equation}
	%\label{eq:Xi}
	%\begin{aligned}
	%\Xi(t) &= \int_0^\infty x e^{-\pi \lambda_\B x^2 \left(\mathbb{E}_f[\Omega(f_b,\delta_\G,t)]+(\lambda_\D/\lambda_\B)\hat P_{\B,\M}^{\delta_\G}(-jt)^{\delta_\G}\right)}dx\\
	%&= \frac{(2\pi\lambda_\B)^{-1}}{\mathbb{E}_f[\Omega(f_b,\delta_\G,t)]+(\lambda_\D/\lambda_\B)\hat P_{\B,\M}^{\delta_\G}(-jt)^{\delta_\G}}.
	%\end{aligned}
	%\end{equation}
	Using $\mathbb{E}_f[\Omega(f_b,\delta_\G,t)]={}_2F_1(\Psi_\G,U_\B;1-\delta_\G;jt)$ in (\ref{eq:cov}), we arrive at (\ref{eq:coverage}). 
	
	\subsection{Proof of Proposition \ref{prop:IoTDistributions}}\label{app:prop1}
	Let the IoT device be at a distance $x_\M$ from the drone. Then, using the mean LOS probability, the received signal power at the drone can be approximated as  $\mathbb{P}(S_\M\leq \tau) \approx \mathbb{P}\big(P_\M L_\M x_\M^{-\alpha_\A} \leq \tau\big)$. W can further simplify this expression to 
	\begin{equation}
	\begin{aligned}
	\mathbb{P}\big(P_\M L_\M x_\M^{-\alpha_\A} \leq \tau\big) %&= \mathbb{P}\big(P_\M L_0 x_\M^{-\alpha_\A} L_0((1-L_{\operatorname{NLOS}}) \mathbb{P}_{\operatorname{LOS}}(x_\M,h_\D) +L_{\operatorname{NLOS}}) \leq \tau\big)\\
	%&\stackrel{(a)}{\approx} \mathbb{P}\big(P_\M L_\M x_\M^{-\alpha_\A} \leq \tau\big)\\
	&= 1-F_{x_\M}\left(\left(\frac{P_\M L_\M}{\tau}\right)^{1/\alpha_{\A}}\right)\\
	&\stackrel{(a)}{=}\left\{\begin{array}{ll}
	1- \frac{(P_\M L_\M/\tau)^{\delta_\A}-h_\D^2}{R^2}, &\frac{P_\M L_\M}{(R^2+h_\D^2)^{1/\delta_\A}}\leq \tau \leq \frac{P_\M L_\M} {h_\D^{1/\delta_\A}}\\
	0,													&\text{otherwise} 
	\end{array}\right.,
	\end{aligned}
	\end{equation}
	where $(a)$ follows using the CDF of the distance between the typical IoT device and the drone $F_{x_\M}(\cdot)$, which is found using the fact that the IoT device is randomly distributed over an area of radius $R$ centered around the 2D coordinates of the drone. Further, let $r_\B$ denote the distance between the tagged BS and the drone, then 
	$\mathbb{P}(I_\B\leq \tau) \approx\mathbb{P}\big(P_\B L_\B r_\B^{-\alpha_\A} \leq \tau\big)
	= 1-F_{r_\B}\left(\left(\frac{P_\B L_\B}{\tau}\right)^{1/\alpha_{\A}}\right)$. 
	Since the distance between a point in $\mathbb{R}^2$ and the nearest BS is distributed as $f_{y_\B}(y)=2\pi\lambda_\B y \exp(-2\pi\lambda_\B y^2)$ \cite{LinLiang2015}. Hence, the distribution of $r_\B=\sqrt{y_\B^2 + h_\D^2}$ can be shown to be  $f_{r_\B}(r) = \frac{2rf_{y_\B}(\sqrt{r^2-h_\D^2})}{\sqrt{r^2-h_\D^2}}$. Thus, we can compute $F_{r_\B}(r)=\int_{h_\D}^{\infty}f_{r_\B}(r)dr$ to arrive at (\ref{eq:BSInterference}).
	
	\subsection{Proof of Proposition \ref{prop:SCSDQCNC}}
	We first derive useful properties for $r(p)$. In particular, $r(p)$ is a non-negative sum of coverage probabilities, each is non-decreasing with the transmit power, and thus $r(p)$ is non-decreasing with $p$. Thus, the $t$-sublevel sets, i.e., $\mathcal{R}_{t,\operatorname{sub}}=\{p|r(p)\leq t\}$, and the $t$-superlevel sets, i.e.,   $\mathcal{R}_{t,\operatorname{sup}}=\{p|r(p)\geq t\}$, are convex, and hence $r(p)$ is quasilinear \cite{BoydVandenberghe2004}. To show that $r(p)$ is S-shaped, we take the 2nd derivative with respect to $p$ to get
	\begin{equation}
	\begin{aligned} 
	\frac{d^2r(p)}{dp^2}&  =\bar c_{\B}\tilde L_\M^2  \sum_{k=1}^{K_\M} \frac{\mu_k e^{-\pi \lambda_\B \left(\frac{\frac{p \tilde L_\M}{\tau_{\M,k}} - P_N}{P_\B L_\B}\right)^{-\delta_\A}}}{\tau_{\M,k}^2  \left(\frac{p \tilde L_\M}{\tau_{\M,k}} - P_N\right)^{2(1+\delta_\A)}}\times \Bigg\{\bar c_{\B}
	- (1+\delta_\A) \left(\frac{p \tilde L_\M}{\tau_{\M,k}} - P_N\right)^{\delta_\A}\Bigg\},
	\end{aligned} 
	\end{equation} 
	where $\bar c_{\B}= \delta_\A \pi  \lambda_\B (P_\B L_\B)^{\delta_\A}$. Note that $\Big(\frac{\frac{p \tilde L_\M}{\tau_{\M,k}} - P_N}{P_\B L_\B}\Big)\geq0$ for any feasible $\tau_{\M,k}$ and the derivative is non-increasing with $p$. Furthermore, there exists a point $p^\bullet$ such that $\frac{d^2r(p)}{dp^2}\geq0$ for $p\leq p^\bullet$ and  $\frac{d^2r(p)}{dp^2}\leq$ for $p\geq p^\bullet$, and hence the function is S-shaped. %Note that when $\tau_k \leq \frac{P^{\operatorname{min}}_\M \tilde L_\M}{P_N+P_\B L_\B \left(\frac{\delta_\A\pi\lambda_\B}{1+\delta_\A}\right)^{1/\delta_A}}~\forall k,$
	%then $r(p)$ becomes concave. Specifically, it is sufficient to show that the coverage $\mathbb{P}(\hat \gamma_\M\geq \tau_k)$ is concave in $p$ as $r(p)$ is a non-negative sum of the coverage probabilities. This can be shown by by taking the second derivative of the coverage and showing that it is non-positive if and only if $\bar c_{\B} \leq (1+\delta_\A) \left(\frac{\frac{p \tilde L_\M}{\tau_k} - P_N}{P_ \B L_\B}\right)^{\delta_\A}$, which is satisfied when the condition on $\tau_k$ holds. 
	
	To prove that the objective function in (\ref{eq:SCSDOpt}) is quasiconcave, then it is sufficient to prove that the superlevel sets $\mathcal{E}_{t,\operatorname{sup}}=\{p|\frac{r(p)}{c(p)}\geq t\}$ are convex. Since the objective is non-negative, then we only consider the case for $t\geq0$ and prove that $\{p|r(p)-c(p)t\geq 0\}$ is convex. 
	We consider two cases. First, for $p\geq p^\bullet$, $r(p)$ is concave, and thus for $p_1,p_2\in\mathcal{E}_{t,\operatorname{sup}}$ and $\beta\in[0,1]$ we have
	\begin{equation}
	\begin{aligned}
	r(\beta p_1 + (1-\beta)p_2)-c(\beta p_1 + (1-\beta)p_2)t 
	&\stackrel{(a)}{\geq} \beta r(p_1) + (1-\beta) r(p_2)-c(\beta p_1 + (1-\beta)p_2)t\\
	&\stackrel{(b)}{=} \beta \left(r(p_1)-c(p_1)t\right) + (1-\beta) \left(r(p_2)- c(p_2)t\right)\\
	&\geq 0,
	\end{aligned}
	\end{equation}
	where $(a)$ follows because $r(p)$ is concave and $(b)$ follows because $c(p)$ is affine, and thus, $\mathcal{E}_{t,\operatorname{sup}}$ is convex. Second, for $p\leq p^\bullet$, $r(p)$ is convex. Taking the first derivative of $\bar E(p)$, we get
	\begin{equation}
	\begin{aligned}
	\frac{d\bar E(p)}{dp}  = \frac{\eta^{-1} (p r'(p)-r(p))+P_{\operatorname{CP}} r'(p)}{c^2(p)}.
	\end{aligned}
	\end{equation}
	The second sum term $r'(p)\geq0$ as $r(p)$ is a non-decreasing function in $p$. In addition, the first sum term $pr'(p)-r(p)\geq0$ because $r(p)$ is convex, i.e., for any convex differentiable function $g(x)$ with $g(0)=0$, we have $g(y)\geq g(x)+(y-x) g'(x) $ \cite{BoydVandenberghe2004}, and by setting $y=0$ we get $x g'(x)\geq g(x)$. Thus, the derivative is non-negative, i.e., it is a non-decreasing function in $p\leq p^\bullet$, and hence the superlevel sets are convex, which completes the proof that the objective function is quasiconcave. To prove that the objective function is unimodal, note that $\bar E(0)=\bar E(\infty)=0$, and since $\bar E(p)$ is quasiconcave, then it has to be unimodal. 
	%---------------------------------------------------------------------------------
	%                         References
	%---------------------------------------------------------------------------------
	\bibliographystyle{IEEEtran}
	\bibliography{C:/Users/ghait/Dropbox/References/IEEEabrv,C:/Users/ghait/Dropbox/References/References}
	
\end{document}